\documentclass[12pt,letterpaper]{article}
\usepackage[a4paper, total={7in, 10in}]{geometry}

\usepackage{graphicx}
\usepackage{helvet}
\usepackage{authblk}
\usepackage{hyperref}
\usepackage{caption}
\usepackage{amsmath} 
\usepackage{mathtools}
\usepackage{amssymb} 
\usepackage{orcidlink} 
\usepackage[super,comma,sort&compress]  
   {natbib}
\usepackage[right]{lineno} \linenumbers
\usepackage[linesnumbered,lined,boxed,commentsnumbered,ruled,longend]{algorithm2e}
\usepackage{xcolor}
\usepackage{textcomp}
\usepackage{pdflscape} 

\makeatletter
\renewcommand{\maketitle}{\bgroup\setlength{\parindent}{0pt}
\begin{flushleft}
  \textbf{\@title}
  
  \@author
\end{flushleft}\egroup}
\makeatother

\title{The Liquid Buffer: Multi-Year Storage for Defossilization and Energy Security under Climate Uncertainty}
\date{}

\author[1,2,*]{Leonard Göke}
\author[3]{Jan Wohland}
\author[1]{Stefano Moret}
\author[1]{André Bardow}

\affil[1]{Energy and Process Systems Engineering, ETH Zurich, 8092 Zurich, Switzerland}
\affil[2]{Reliability and Risk Engineering, ETH Zurich, 8092 Zurich, Switzerland}
\affil[3]{Department of Technology Systems, University of Oslo, 2007 Kjeller, Norway}

\affil[*]{Correspondence: lgoeke@ethz.ch}

\begin{document}

\maketitle

\section*{Abstract}

The climate-driven uncertainty of renewable generation and electricity demand challenges energy security in net-zero energy systems. By introducing a scalable stochastic model that implicitly accounts for 51'840 climate years, this paper identifies multi-year storage of liquid hydrocarbons as a key option for managing climate uncertainty and ensuring energy security. In Europe, multi-year storage reduces system costs by 4.1\%, fossil imports by 86\%, and curtailment by 60\%. 

The benefit of multi-year storage is that a renewable surplus in one year is not curtailed but converted to synthetic oil, with hydrogen as an intermediate product, and stored to balance a future deficit. We find that the required energy capacity for liquid hydrocarbons is 525 TWh, a quarter of the European Union's current oil and gas reserves, complemented by 116 TWh for hydrogen storage. Security of supply remains high and unserved energy only amounts to 0.0035\textperthousand\, well below the common target of 0.02\textperthousand.

\section*{Introduction}

Mitigating climate change requires climate-dependent energy sources, like wind, solar, and hydro, to supply the power sector and electrify heating, transport, and industry. Climate-dependent supply adds to the already climate-sensitive demand and thus to the challenges of future energy systems. Over short periods, supply and demand mismatches are manageable thanks to reliable weather forecasts and technical solutions, like short-term storage, particularly batteries, demand-side flexibility, and power grids \citep{Goeke2023,Brown2018}. However, the longer the mismatches are, the harder they become to manage.

Last winter, European power systems were already under stress when the wind generation was low, and cold temperatures induced high demand, causing a multi-day deficit, also referred to as Dunkelflaute \citep{wiel2019,cew2024,Kittel2024}. In the future, the expansion of renewables and electrification will aggravate the problem, extending the deficit from a few days to multiple months or even years \citep{Most2024,Jung2018}. Prolonged deficits can result from years with high heating demand but low generation from wind and hydropower \citep{Wohland2021,Gotske2024}. For instance, South America already faced multi-year energy droughts due to subpar rainfalls cutting vital generation from hydropower \citep{nyt2024}. Initially, hydro storage backed up supply, but once depleted, countries faced a severe energy crisis and scheduled blackouts. In Africa, multi-year droughts caused energy shortages and blackouts in Zambia and Zimbabwe at the same time \citep{guardian2024}.
 
\begin{figure}[h]
 \centering
  \includegraphics[scale=0.65]{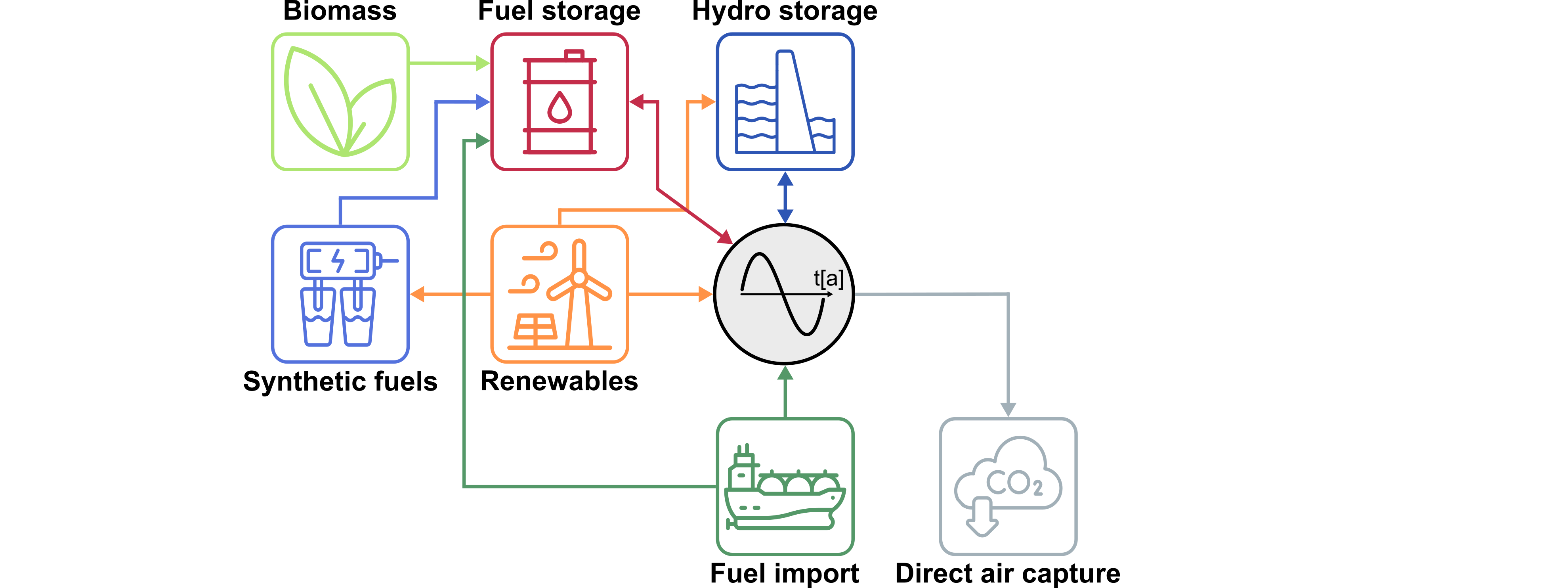}
 \caption[LoF entry]{Overview of security options that can manage mismatches of supply and demand lasting longer than multiple days. Incoming arrows to the middle circle indicate additional supply; outgoing edges indicate additional demand. Consequently, storages have two-sided edges. Other edges represent independence between options.}
 \label{fig:secOpt}
\end{figure}

Several security options (Fig. \ref{fig:secOpt}) can manage mismatches of supply and demand that last longer than multiple days, which is too long for battery storage or demand-side management \citep{Schmidt2023}. One security option is to overbuild renewable capacity to ensure sufficient supply even during energy droughts. However, during non-drought periods, overcapacity will result in massive curtailment. As an alternative to overcapacity, storage can shift renewable supply from surplus to drought periods. Hydro storage achieves the highest round-trip efficiency, but its technical potential is limited; storing energy as synthetic fuels is easier at a large scale, but producing these fuels requires electricity or biomass, whose availability is limited.

Fuel imports are another option to balance supply and demand in the long term, particularly flexible imports during energy droughts. However, the European energy crisis showed that fuel imports as a backup must be treated cautiously: First, a failure of fuel imports caused the crisis, and second, it revealed that in the short-term, the international market could not reliably compensate for the deficit \citep{reuters2022}. Market liquidity is limited since most deliveries are tied to long-term contracts and require dedicated infrastructure, like pipelines or terminals \citep{iea2025}. In the case of renewable fuels, exporting countries can also not flexibly increase the climate-dependent production.

Finally, direct air capture can add flexibility if the system does not need to comply with a greenhouse gas emission target each year, but only in the long-term average. In this case, negative emissions from direct air capture during surplus periods compensate for positive emissions from fuel consumption during drought periods.

Evaluating all these security options is a stochastic problem since climate conditions are uncertain and cannot be forecasted months or years into the future. Consequently, the risk of an energy drought must be weighed against the costs of security options to prevent overengineered systems, hedged against virtually impossible droughts. Particularly, storage across multiple years under uncertainty is pivotal for the analysis, yet complex to model due to interdependencies over extensive time frames.

Furthermore, an analysis of climate security must take a system perspective to capture the interactions between security options and the climate sensitivity of supply and demand. For instance, more PV will reduce supply variance across years while increasing seasonal imbalances since generation in winter is low \citep{Lledo2022}. In the heat sector, electrification reduces total demand but increases variance during winter since the efficiency of heat pumps is temperature-sensitive. 

The existing literature only considers energy droughts lasting up to one year and is consequently confined to seasonal storage, omitting inter-annual variability \citep{Ghanbarzadeh2025}. However, seasonal and multi-year droughts and storage are mutually dependent \citep{Most2024}. In addition, previous studies on energy droughts are deterministic, only assuming climate as \textit{variable} but not as \textit{uncertain}, hence they consider how supply and demand vary due to the weather, but assume these variations are perfectly known, even months into the future \citep{Schmidt2025}. Yet, due to the chaotic nature of atmospheric physics, it is impossible to make such conclusive long-term forecasts \citep{Lorenz1969}.

Specifically, the existing literature on energy droughts uses two approaches: One approach runs more detailed system models separately for different climate years, assuming perfect foresight of climate conditions in each year 
\citep{Grochowicz2023,Grochowicz2024,Gotske2024,DeMarco2025}. Since this approach only considers single years in isolation, it cannot provide insights on multi-year droughts or storage. A second approach runs a single stylized model with one multi-year time series of up to 40 years \citep{Zeyringer2018,Dowling2020,Ruhnau2022,Brown2023,Ruggles2024}. However, these models have perfect foresight over the entire time horizon and, as a result, will anticipate droughts decades in advance---contrary to the uncertainty of climate that prevents reliable forecasts. In addition, computational limits restrict the multi-year approach to stylized models of a single region and only the power sector. Yet, in the future, energy droughts will arise in highly interconnected and integrated energy systems. 

In summary, the literature on energy droughts is biased due to assuming perfect foresight and lacks a multi-year perspective. These gaps result from using deterministic methods limited to climate \textit{variability} to solve the stochastic problem of climate \textit{uncertainty}. In this light, review articles recommend stochastic methods to study climate uncertainty in integrated energy systems but also acknowledge the computational difficulty \citep{Craig2022,Plaga2023,DossGollin2023}. As a notable first step, \citet{Schmidt2025} proposes a stochastic approach to climate uncertainty based on stochastic dual dynamic programming confined to seasonal storage and a stylized single-region model.

This paper demonstrates how to secure net-zero energy systems against climate uncertainty. We focus on the benefits of multi-year storage, comparing multi-year to seasonal storage and flexible imports. Since the reliability of imports is disputable, additional scenarios investigate security when limiting the flexibility of imports or excluding them entirely.

For the study, we introduce a stochastic but scalable approach for energy modeling under climate uncertainty, assuming, based on autocorrelation analysis in previous work, that foresight and stochastic independence of climate conditions are limited to one month \citep{Schmidt2025}. On this basis, we develop a robust model that implicitly considers up to 51'840 hourly resolved climate years for system planning, but explicitly only models 32 representative months. For the selection of representative months, we propose an optimization-based clustering algorithm. 

We apply the stochastic approach to a case study on the integrated European energy system. The European model covers the power, heat, transport, and industry sectors to account for cross-sectoral strategies for adapting to climate uncertainty. To capture the full impact of fluctuating renewables on the security of supply, the model applies a full hourly resolution. We deploy a distributed solution algorithm to solve the extensive stochastic planning problem. Thanks to this algorithm, this study is the first to apply stochastic optimization to a detailed and large-scale problem of macro-energy system planning. The method section provides a comprehensive description of the stochastic approach, the clustering algorithm, and the applied model. All code and data are openly available.

\section*{Results} 

\subsection*{Multi-year storage reduces system costs and diminishes dependence on fossil imports}

\begin{figure}[htbp]
 \centering
  \includegraphics[scale=0.65]{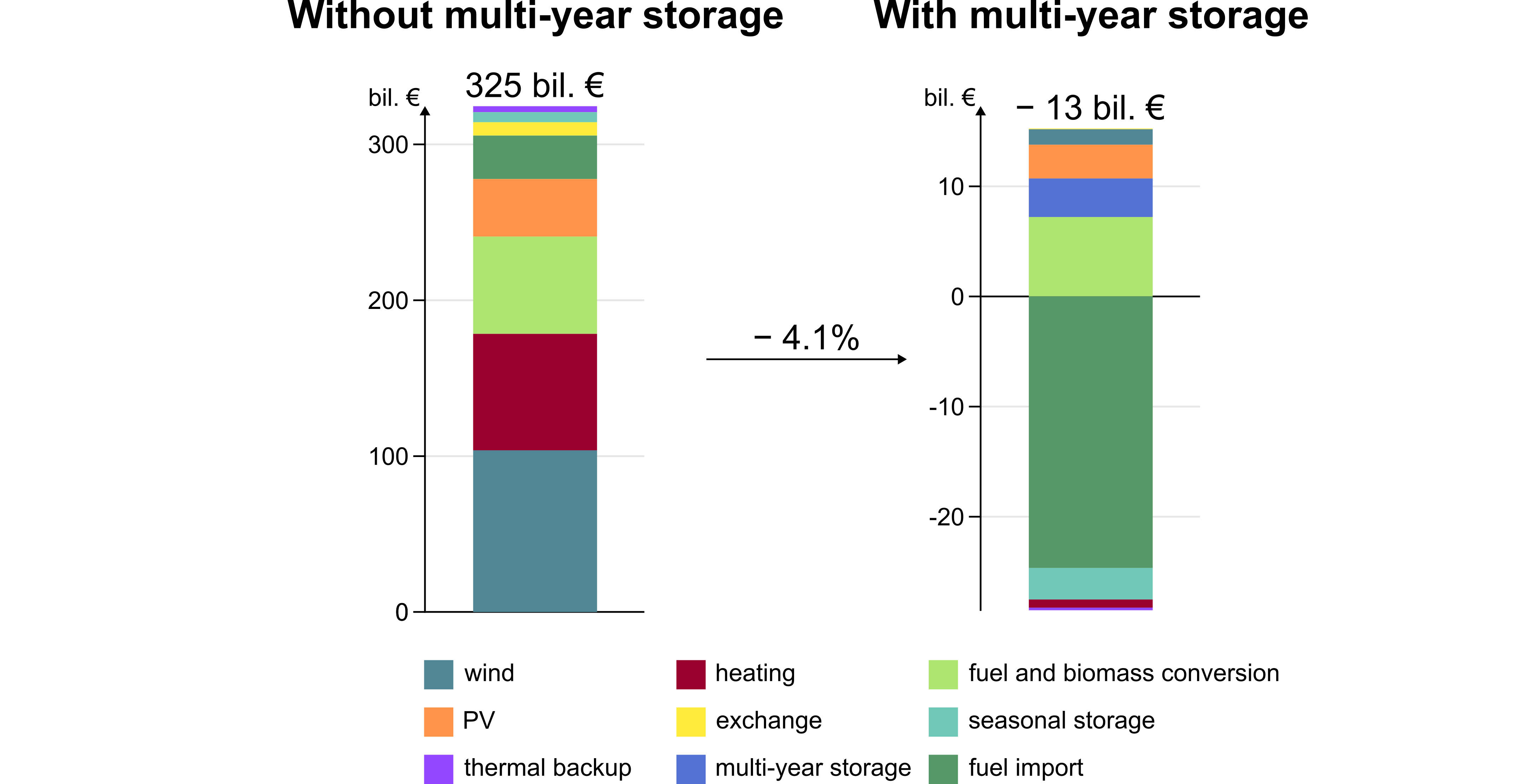}
 \caption[LoF entry]{The bar charts show how multi-year storage reduces system costs in the reference case. The left chart breaks down the system costs without multi-year storage. The chart on the right shows how costs change if multi-year storage is an option.}
 \label{fig:costOver}
\end{figure}

Multi-year storage significantly reduces the costs of energy security in the European energy system. As shown in Fig. \ref{fig:costOver}, multi-year storage reduces system costs by 4.1\% or 13 billion euros per year. This reduction is considered substantial as previous studies found that a carbon network reduces system costs by 3.1\%; a hydrogen grid by 4.7\% \citep{Hofmann2025}. Thus, multi-year storage can be expected to be of similar importance.

The reduction is primarily due to a decline in costs of fossil imports by 24.5 billion, thanks to synthetic fuels from electricity replacing fossil fuel imports. The costs of oil imports decrease by 21.6 billion euros since oil imports drop from 486 to 51 TWh per year. Costs of gas imports drop by 2.9 billion euros; imports drop from 111 to 24 TWh.

The savings in fossil imports are partially offset by increased costs for the production of synthetic fuels. Investments in fuel conversion increase by 7.2 billion euros; to provide the input electricity, investments in wind and solar increase by 4.5 billion euros. Heating costs decrease by one billion, since air-source heat pumps replace more capital-intensive and energy-efficient ground-source heat pumps. This shift occurs since multi-year storage facilitates supplying electricity during cold winters. Finally, increased costs for multi-year storage slightly exceed reduced costs for seasonal storage by 0.3 billion euros. 

Fuel imports can substitute multi-year storage and secure the energy supply. Besides the geopolitical risk, whether relying on imports is beneficial critically depends on their flexibility. The reference case with and without multi-year storage in Fig. \ref{fig:costOver} assumes imports have some flexibility and can increase by up to 10\% at 20\% higher prices based on typical take-or-pay clauses and volume tolerances in fossil supply contracts \citep{Ason2022,Creti2004}.

\begin{figure}[htbp]
 \centering
  \includegraphics[scale=0.65]{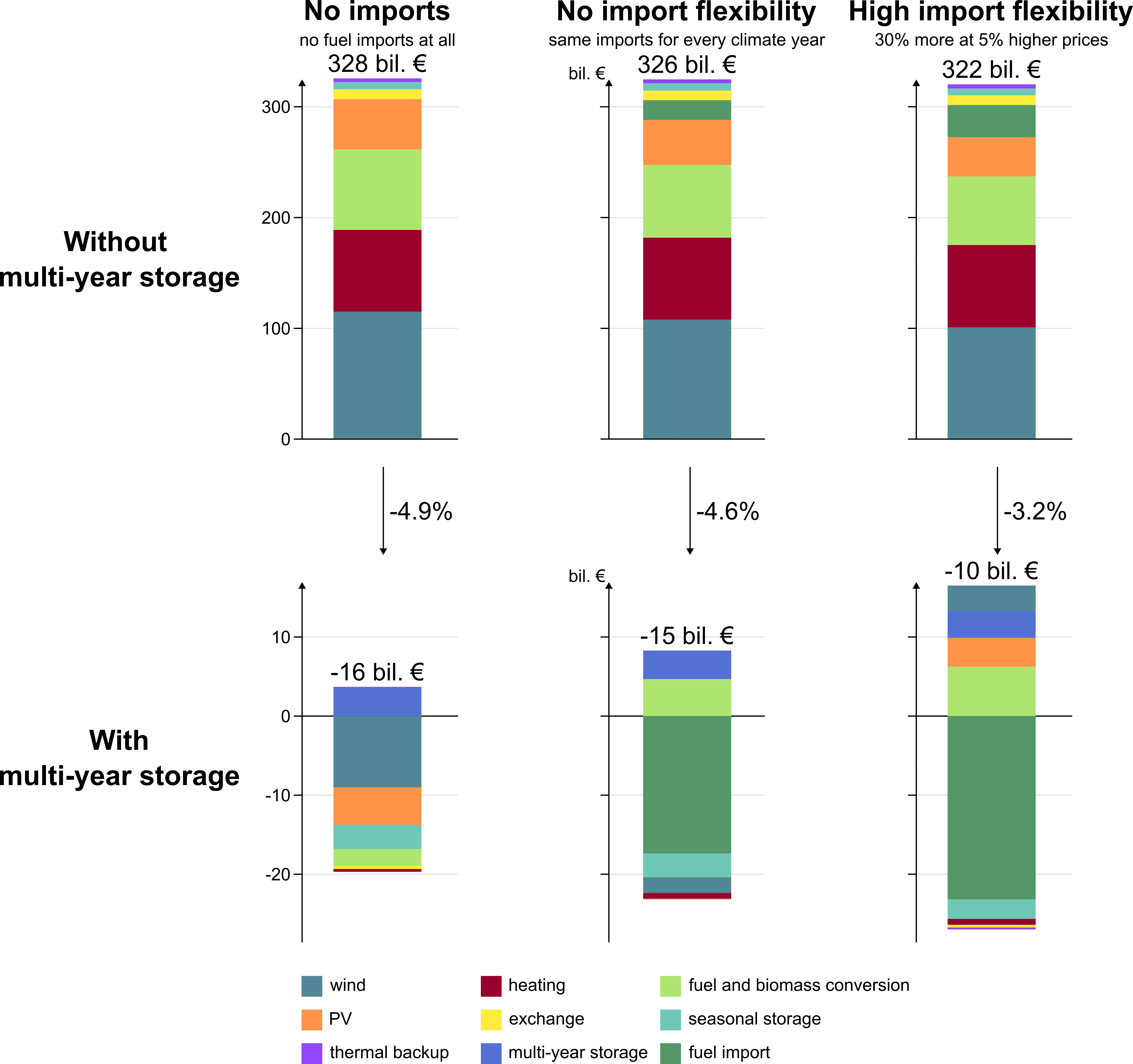}
 \caption[LoF entry]{Cost reductions from multi-year storage depend on the flexibility of imports. The bar charts in the top row show the composition of system costs without multi-year storage for different sensitivities; the bottom row how costs change with multi-year storage.}
 \label{fig:costSensi}
\end{figure}

Against this background, three sensitivities (Fig. \ref{fig:costSensi}) vary the assumed flexibility of fuel imports: The first sensitivity assumes imports are entirely prohibited. As a result, the cost reduction from multi-year storage increases to 4.9\%. For this sensitivity, cost reductions do not originate from replacing fuel imports with renewables, but rather from the reduced curtailment, which decreases investment costs for renewables by 14 billion euros.

In the second sensitivity, imports are inflexible, viz., the same for every climate year. As a result, the cost reduction from multi-year storage increases to 4.6\%. With 17 billion euros, the greatest share of cost reductions still results from replacing fuel imports. Investment in renewables drops by 1 billion since the reduction of curtailment from multi-year storage overcompensates the additional electricity demand for synthetic fuels.

Lastly, if imports are more flexible and can increase up to 30\% at 5\% higher prices, reflecting the upper end of take-or-pay clauses and volume tolerances in fossil supply contracts, the cost reduction from multi-year storage decreases to 3.2\%. Effects are similar to the reference case, but their magnitude is slightly smaller.

\subsection*{Benefit of multi-year storage lies in shifting hydrocarbons from surplus to deficit years}

\begin{figure}[htbp]
 \centering
  \includegraphics[scale=0.65]{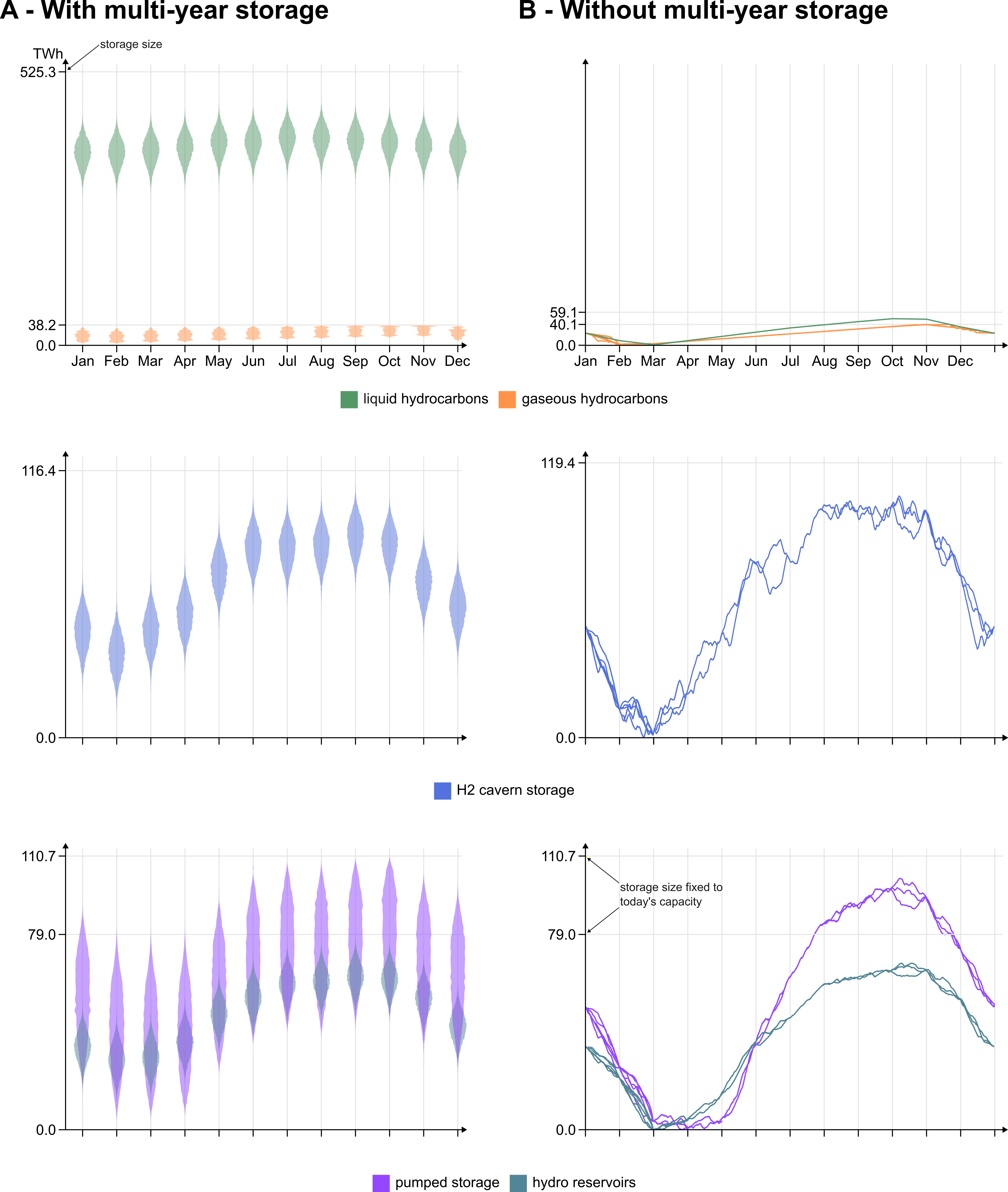}
 \caption[LoF entry]{The graphs show the levels of long-term storage with multi-year storage in the left column and without multi-year storage in the right column. The violin plot for multi-year storage shows how levels vary from year to year. The storage levels were simulated 200 times for 100 years to generate the distributions. Each year in the simulation performs a probability weighted random walk through the set of representative months shown in Fig. \ref{fig:clusterResult} of the method section. The displayed values on the y-axis give the total energy capacity. Since the plot aggregates across regions, aggregated levels only ever reach the total energy capacity if all regions reach their maximum level simultaneously.}
 \label{fig:storageLevels}
\end{figure}

The storage levels in Fig. \ref{fig:storageLevels} explain how multi-year storage enables the observed cost savings. Since levels of multi-year storage vary not only by month but also by year, the graphs on the left show monthly distributions rather than a single curve.

Storage of fossil and synthetic oil, viz., liquid hydrocarbons, constitutes the largest share of multi-year storage. Compared to the case without multi-year storage, energy capacity for liquid hydrocarbons increases from 59 to 525 TWh. 95\% of the 525 TWh are dedicated to multi-year storage, reflected by an almost nonexistent seasonal profile for liquid hydrocarbons in the case with multi-year storage. The energy capacity of gas storage decreases slightly to 38 TWh with multi-year storage; multi-year storage has a share of 54\%.

Hydrogen storage, pumped storage, and hydro reservoirs contribute to multi-year storage to a lesser extent, as indicated by their pronounced seasonal patterns, similar to the case without multi-year storage. Pumped storage refers to storage that can be actively charged with surplus electricity, whereas hydro reservoirs have an exogenous and climate-dependent inflow. Of the three technologies, a total energy capacity of 170 TWh (55\%) is dedicated to multi-year storage. For hydrogen storage, energy capacity increases slightly by 3 TWh if multi-year storage is an option; multi-year storage has a share of 46\%. For pumped storage and hydro reservoirs, total capacity is fixed and thus remains constant; the share of multi-year storage is 61\%.

\begin{figure}[htbp]
 \centering
  \includegraphics[scale=0.7]{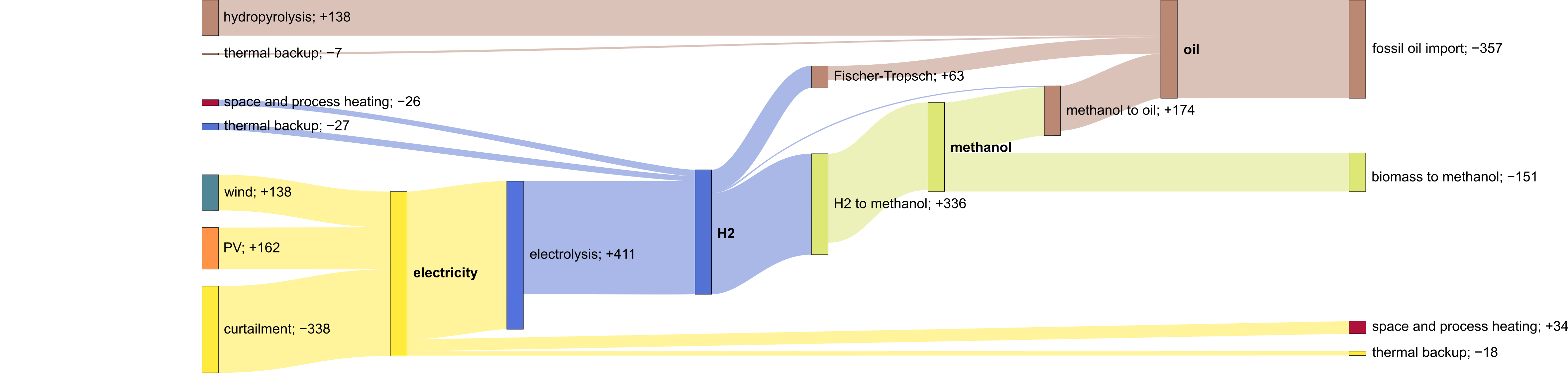}
 \caption[LoF entry]{The diagram shows how yearly energy flows in TWh change with multi-year storage. The bold nodes represent energy carriers, and the other nodes represent technologies. Technologies with increasing generation or decreasing demand enters carrier nodes from the left; vice versa, decreasing generation, or increasing demand exit on the right. All values are expected values across the 51'840 considered climate years. For clarity, the figure aggregates and simplifies the actual flows in the underlying model. For a detailed model description, see the method section. Furthermore, the appendix provides more detailed Sankey diagrams with absolute values.}
 \label{fig:sanDelta}
\end{figure}

The change of energy flows in Fig. \ref{fig:sanDelta} shows how multi-year storage of liquid hydrocarbons serves as a buffer that enables substituting fossil imports with renewables. With multi-year storage, a renewable surplus in one year is converted to synthetic oil, stored, and used to meet demand when an energy deficit occurs. Without this option, the system curtails renewable surpluses and imports more fossil fuels.

With multi-year storage, fossil oil imports drop by 357 TWh. To compensate, synthetic oil production from methanol increases by 174 TWh; the Fischer-Tropsch process produces 63 TWh using carbon from direct-air capture. Hydrogen production from electrolysis increases by 411 TWh to provide the inputs for synthetic oil production. Besides reducing curtailment by 338 TWh, renewable electricity generation increases to produce the required hydrogen. With multi-year storage, wind generation increases by 138 TWh (2.8\%); solar generation by 162 TWh (8.1\%).

In addition, multi-year storage facilitates electricity supply during deficit periods, shifting the supply of space and process heat from hydrogen to electricity, and reducing the need for thermal backup plants. Finally, a shift in the processing routes of methanol and oil occurs, as we assume there is no option for multi-year storage of methanol, only for oil. As a result, oil production from hydropyrolysis increases by 138 TWh, while methanol production from biomass decreases by 151 TWh.

\begin{figure}[htbp]
 \centering
  \includegraphics[scale=0.65]{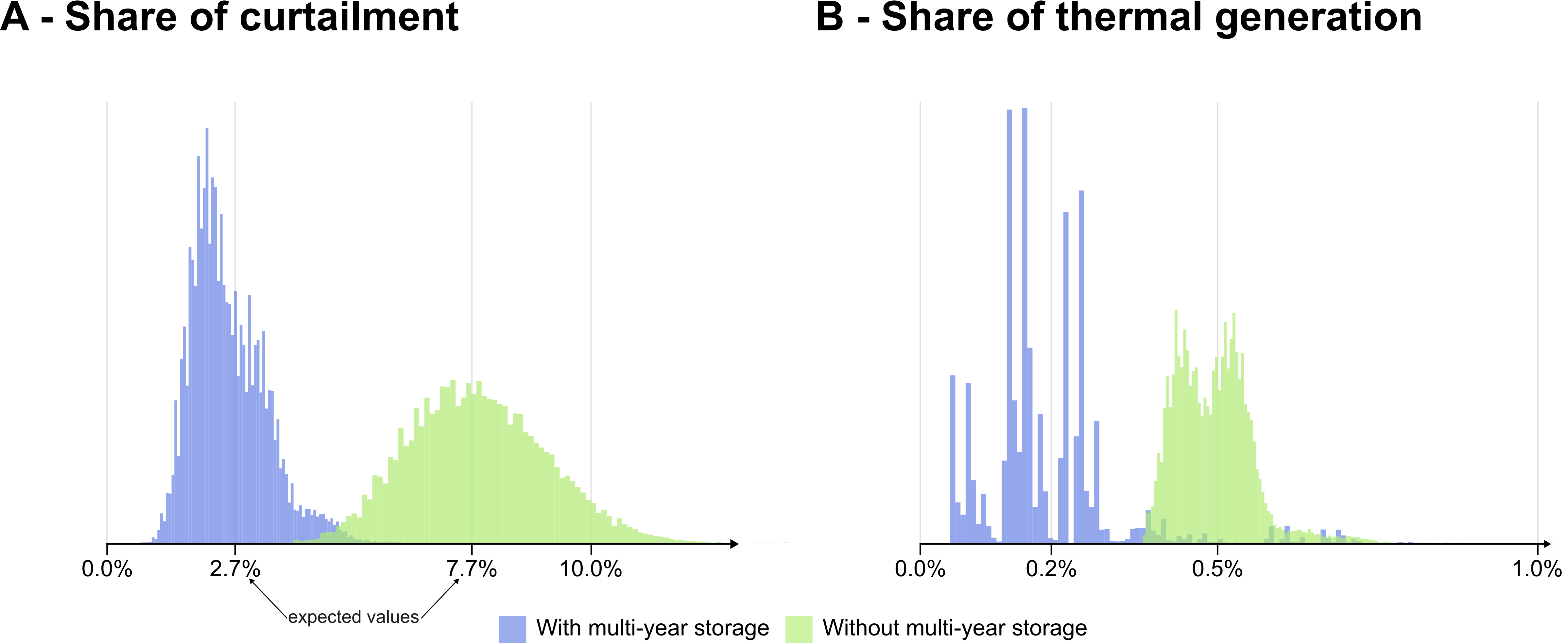}
 \caption[LoF entry]{The figures show the probability density function for curtailment relative to expected renewable generation and thermal generation relative to the expected total generation. The distributions result from computing curtailment and thermal generation for the 51'840 considered climate years.}
 \label{fig:dist}
\end{figure}

The probability density functions of curtailment and thermal generation in Fig. \ref{fig:dist} illustrate how multi-year storage serves as a long-term flexibility option that integrates renewables by reducing curtailment and the need for thermal backup generation. With multi-year storage, the expected share of curtailment decreases by 5\% to 2.7\%. The expected share of thermal generation more than halves from 0.48\% to 0.22\%.

\subsection*{Multi-year storage balances energy surpluses and deficits driven by wind supply during winter}

\begin{figure}[htbp]
 \centering
  \includegraphics[scale=0.65]{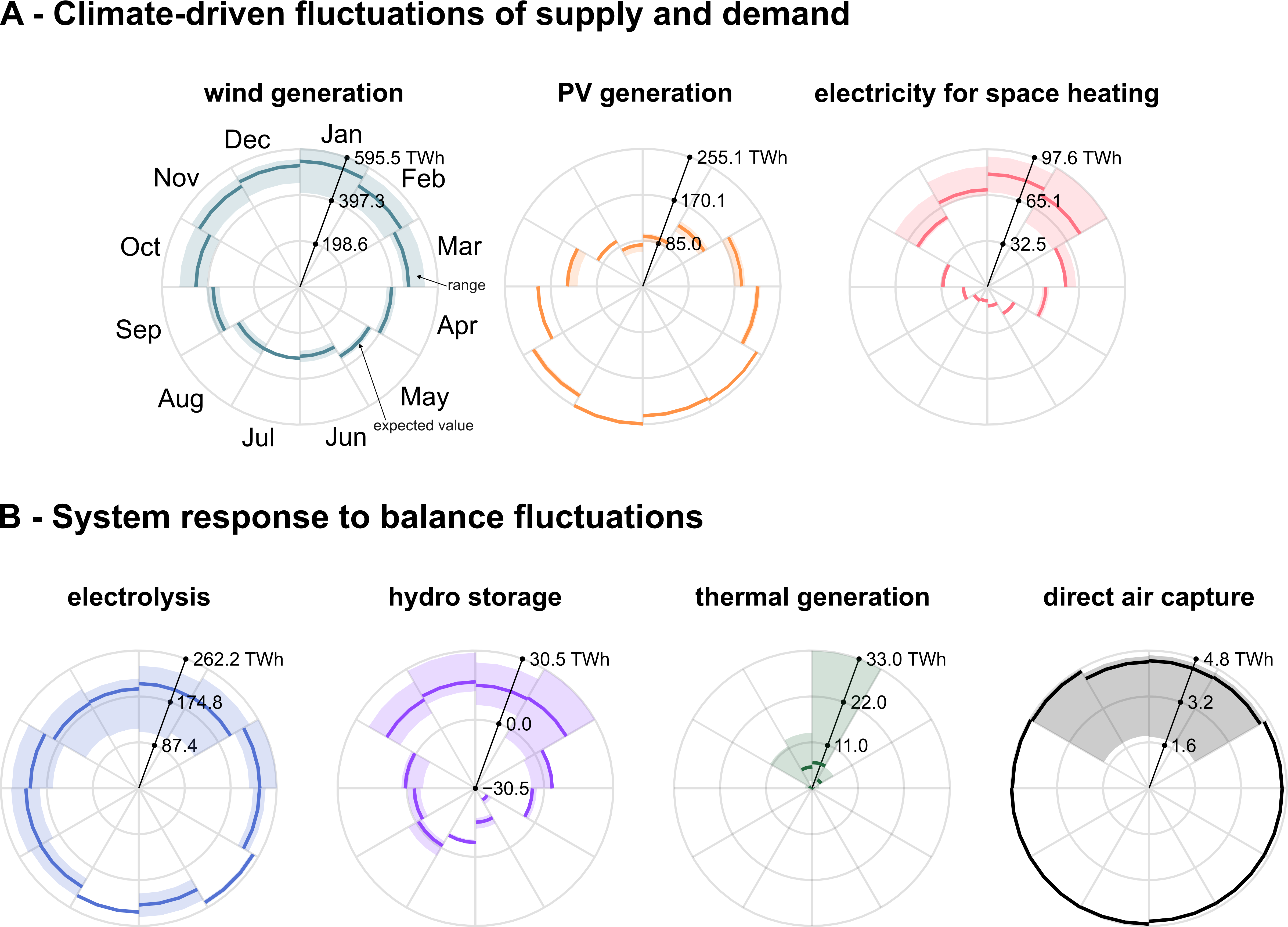}
 \caption[LoF entry]{The radar plots show the climate-driven range of supply and demand (A) and the corresponding system response (B). The shaded areas represent the range of values for each month; the line represents the monthly average. All values refer to electricity supply or demand.}
 \label{fig:circles}
\end{figure}

The seasonal patterns and total range of supply and demand (part (A) of Fig. \ref{fig:circles}) reflect the variability and uncertainty of supply and demand that the system---enabled by multi-year storage---reacts to (part (B) of Fig. \ref{fig:circles}).

Wind generation accounts for the largest share of climate uncertainty; PV and electrified heating for a much smaller share. For wind, expected generation and the uncertainty range peak in winter, reaching an expected value of 540 TWh with a 191 TWh uncertainty range in January. The 95\% confidence interval of yearly generation spans from 4883 to 5368 TWh.

Solar generation exhibits significant seasonal variability, inverse to wind, but little uncertainty. The 95\% confidence interval of yearly generation only spans from 2139 to 2168 TWh. Electricity demand for heating only occurs during winter, but its uncertainty is also small. The yearly confidence interval spans from 613 to 669 TWh. These results consider the effect of temperature on heating demand and heat pump efficiency. 

Generally, the system adapts to climate conditions by increasing supply and reducing demand during winter (part (B) of Fig. \ref{fig:circles}). Analogously to the findings in part (A), the uncertainty of supply and demand is largest in winter. Multi-year storage enables this year-to-year flexibility, viz., the deviation from a fixed seasonal profile in the graphs. For instance, hydrogen production can vary, as above-average production is converted into synthetic oil and stored, compensating for below-average production in another year.

Adapting hydrogen production from electrolysis is crucial for balancing an energy surplus or deficit. Analogously to wind generation, the uncertainty range of electricity consumption peaks in January at 125 TWh. The 95\% confidence interval for yearly electricity consumption from electrolysis ranges from 2430 to 2726 TWh. 

Hydro storage and thermal generation also adapt to climate conditions, but the extent is much smaller. Exogenous (and climate-dependent) inflows constrain hydro storage and limit the adaptation to supply. From November to February, the average net-discharge remains almost constant at around 13 TWh. Thermal generation is limited to deficit months in winter due to high operating costs, but low investment costs. The average monthly generation from November to February amounts to only 4 TWh; however, in the most extreme case, generation peaks at 33 TWh in a high-deficit January. In total, thermal plants have expected full load hours of 230 hours, and their electrical capacity is 133 GW with 60\% of capacity from diesel engines.

Direct air capture only reduces its electricity consumption in rare winter months with an extreme deficit. More flexible operation would require additional capacity, which is not economical, since investment costs are high, particularly compared to the more flexibly operated electrolyzers.

The system rarely cuts demand as a last resort for managing critical deficits; thus, high energy security is high. The expected normalized energy not served is at 0.0035\textperthousand, well below the common contemporary threshold of 0.01\textperthousand to 0.02\textperthousand.\citep{usde}. Since deficit years are also met with increased thermal generation and reduced direct air capture, energy not served and emissions are closely correlated. In most years, emissions are slightly negative, compensating for the significant positive emissions during the few deficit years.

\section*{Discussion}

Multi-year storage is shown to be a cost-efficient option for securing a net-zero energy system against climate uncertainty. Serving as an energy buffer, multi-year storage reduces system costs and the dependence on fossil fuels while supporting the integration of renewable energy. Although renewables and electrification make a net-zero energy system more sensitive to climate conditions, our results show that systems can still achieve high supply security. Not enforcing net-zero for every single year, but as an average over multiple years, is beneficial for system security. In this case, positive emissions reduce stress in years with an extreme energy deficit, but require negative emissions in the other years.

Most relevant is the multi-year storage of liquid hydrocarbons. With this option, a climate-driven surplus of PV and wind in one year is not curtailed but converted to synthetic oil, with hydrogen as an intermediate product, and stored to balance a future deficit. The required energy capacity to achieve the above benefits amounts to 525 TWh, only a quarter of the European Union's current oil and gas reserves \cite{EU1,EU2}. Hydrogen and hydro storage are relevant too, but primarily operate as seasonal storage.

Our results substantially deviate from previous studies---qualitatively and quantitatively. Most importantly, previous studies only consider climate variability, not uncertainty; thus, they make the unrealistic assumption of perfect forecasts far into the future. As a result, previous work on long-term climate impacts on European net-zero systems exclusively suggested hydrogen instead of liquid hydrocarbons for long-term storage; energy capacities from 15 to 30 TWh, compared to a total of 680 TWh in our study \citep{Grochowicz2024,Gotske2024,DeMarco2025}. Other studies have proposed liquid hydrocarbons for long-term storage, specifically methanol. However, these works focused on technical advantages over hydrogen and did not quantify the storage needs for energy security against climate uncertainty \citep{Brown2023,Glaum2025}.

This study proposes a climate-secure energy system utilizing long-term storage. However, our method cannot provide a strategy to coordinate the storage during operation, prescribing, for instance, if surplus electricity should charge the hydro storage or be converted to fuels and stored then. \citet{Schmidt2025} provided first research on this complementary question.

Furthermore, our model of climate uncertainty refrains from assuming perfect weather forecasts and limits foresight to a single month instead. Although a major improvement compared to the status quo, limited foresight still neglects the continuous update of weather forecasts and the very limited but possible prediction of decadal climate trends \citep{Hutchins2025}. Therefore, future work could incorporate the impact of multi-decadal trends, such as the effect of the North Atlantic oscillation on wind generation \citep{Wohland2019}. Additionally, we rely on historical climate data rather than future climate projections, as existing projections lack data on the heating sector. However, previous studies comparing historical and future data do not suggest a major difference \citep{Hou2021,Wohland2025}.

Finally, considering the uncertainty of biomass supply, the multi-year storage of methanol, and the costs of carbon storage will further increase the benefits of multi-year storage, but was infeasible in this study due to computational and data limitations \citep{Gernaat2021}.

In conclusion, our study demonstrates that the uncertainty of climate conditions is manageable and not a dealbreaker for the energy transition. Implementing the energy transition should consider multi-year storage, particularly of liquid hydrocarbons, to reduce costs and enhance energy security by reducing the dependence on fossil fuel imports.

\newpage

\section*{Methods}

The first section describes the energy system model, including the mathematical formulation, the specific setup, and the specific method for the flexibility of fuel imports. The following section introduces the stochastic approach, refining the formulation of the energy system model. The last two sections report on the clustering algorithm to select representative months and the solution algorithm, respectively.

\subsection*{Energy planning model}

This section describes the energy system model's mathematical formulation and setup. In addition, it introduces the implementation of fixed supply contracts for fuel imports with limited flexibility. The model refines and greatly extends a pre-existing model of the European energy system deployed in previous research \citep{Goeke2023,Goeke2025}.

\subsubsection*{Key characteristics}

\begin{figure}[h]
 \centering
  \includegraphics[scale=0.65]{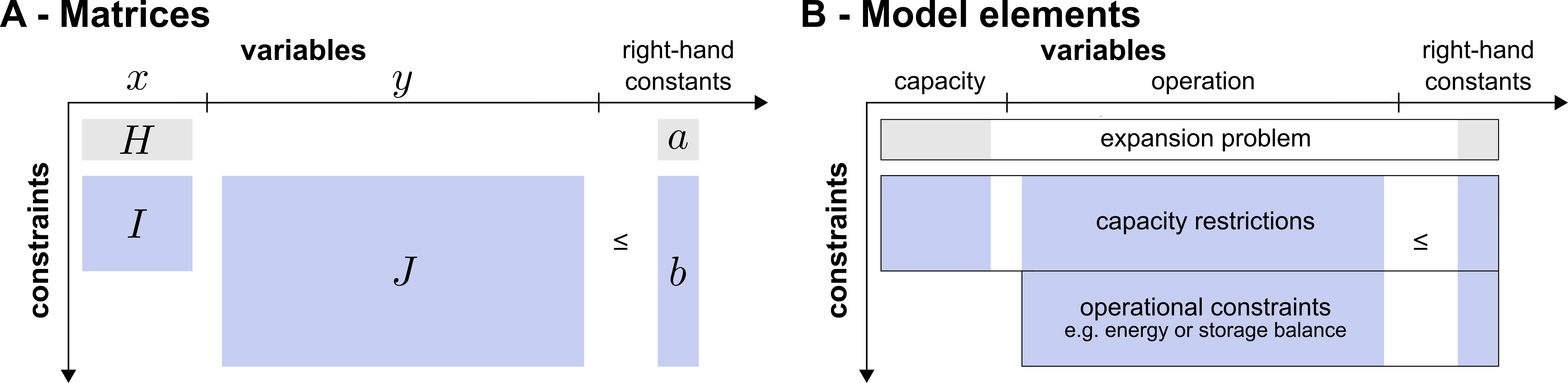}
 \caption[LoF entry]{The matrix form of an optimization problem showing variables and constants on the x-axis and constraints on the y-axis. Part (A) places the coefficient and constant matrices from Eq. \ref{eq:1a} to \ref{eq:1d} in the matrix to illustrate the problem structure. Part (B) groups the matrices, splitting the problem into expansion in the first stage and operation in the second. Note that the figure is not true to size, and the second-stage operation in blue greatly exceeds the expansion problem in grey in terms of variables and constraints. 
}
 \label{fig:matrix}
\end{figure}

The applied planning model is a linear optimization problem of the structure in Eq. \ref{eq:1a} to \ref{eq:1d}. In the first stage, the model decides on $x$, the expansion of capacities; in the second stage on $y$, the operation of capacities. The objective function in Eq. \ref{eq:1a} minimizes the sum of investment and operational costs, defined by the $x$ and $d$, respectively. Eq. \ref{eq:1b} constrains the capacity expansion, for instance, imposing an upper limit on the capacity of wind turbines reflecting available sites. Eq. \ref{eq:1c} constrains the operation. The capacity constraints limit operation to the existing capacities and, thus, include capacity and operational variables, as illustrated in Fig. \ref{fig:matrix}. Any other operational constraints, like the energy or storage balance, only include operational variables. Note that operational variables and constraints scale with temporal detail and greatly exceed capacity variables and constraints, especially in models with high temporal resolution. 

\begin{subequations} 
\begin{alignat}{3}
\min_{x,y} \; \; & c^\top x  \; \: + && d^\top y \label{eq:1a} \\
s.t. \; \; & H x && \leq a \label{eq:1b} \\  
& I x + J y && \leq b \label{eq:1c} \\  
\noalign{\centering $x \in \mathbb{R}^{n}_+,\, y \in \mathbb{R}^{n}_+$} \label{eq:1d}
\end{alignat}
\end{subequations}

\begin{figure}[h]
 \centering
  \includegraphics[scale=0.65]{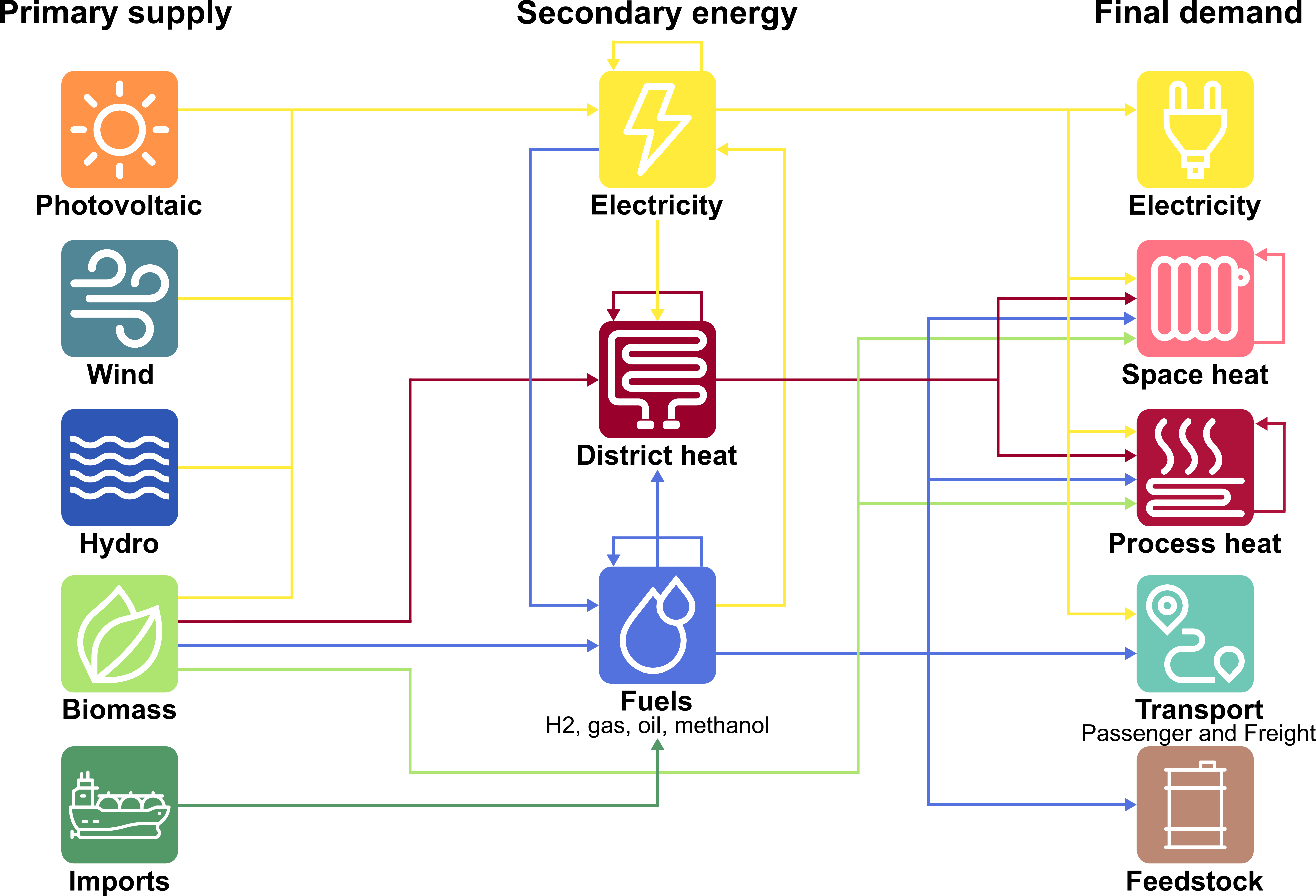}
 \caption[LoF entry]{
The figure illustrates the paths from primary to secondary energy and final demand in the applied energy system model. In the model, the decision on how to meet final demand translates into capacity and operational decisions for 101 conversion, nine storage, and four exchange technologies.}
 \label{fig:overviewEuSysMOD}
\end{figure}

Fig. \ref{fig:overviewEuSysMOD} gives an overview of the considered sources for primary energy, the intermediate secondary energy carriers, and the final demands in the model. For primary energy in the left column, the model can expand capacities for fluctuating renewables, like openspace PV, rooftop PV, onshore wind, and offshore wind. Regarding hydropower, the model considers run-of-river, pumped storage, and reservoirs. Run-of-river fluctuates, like PV and wind. Pumped storage and reservoirs have exogenous inflows, but generation is flexible, given sufficient storage levels. Pumped storage can also actively charge energy. Today's hydro capacities are available without investment, but further expansion is prohibited since the remaining potential of hydro is small and its technical lifetime is long. The primary supply from biomass has a fixed energy limit for each region. The model can utilize this potential by expanding capacities to produce biofuels, heat, or electricity. The model considers hydrogen, synthetic gas and oil, and fossil gas and oil for fuel imports from outside Europe. Emissions from gas and oil require compensation from direct air capture (DAC), which incurs investment costs and requires electricity and process heat.

Using primary energy, the model can generate the secondary energy carriers in the middle column of Fig. \ref{fig:overviewEuSysMOD}, electricity, district heat, and fuels like hydrogen, synthetic gas, or synthetic oil. 
In addition, several technologies can convert one secondary energy carrier into another. For instance, electrolyzers generate hydrogen from electricity and, vice versa, open cycle plants generate electricity, and potentially district heat, from hydrogen.

Utilizing primary and secondary energy, the model satisfies an exogenous final demand for the carriers in the right column of Fig. \ref{fig:overviewEuSysMOD}. The final electricity demand includes residential, service, and industry appliances not used for space heating, process heating, or transport since the model considers these demands separately. Process heat is subdivided into three temperature levels: below 100°C, 100°C to 500°C, and above 500°C; transport is subdivided into six categories: private road, public road, passenger rail, heavy freight road, light freight road, and freight rail. In addition, the model includes an exogenous fuel demand covering aviation, navigation, and feedstocks in the industry.  

\subsubsection*{Technological, temporal, and spatial setup}

The model includes 101 conversion technologies to describe all these conversion pathways in detail. They can be grouped as follows: 
\begin{itemize}
 	\item  \textbf{Electricity generation}. There are 36 technologies for electricity generation, including five technologies for PV: rooftop systems on residential buildings, rooftop systems on industrial buildings with good and poor site quality, and openspace systems with good and poor site quality. Similarly, the model splits offshore wind into good and poor site quality. For PV, site quality affects the capacity factors; for offshore wind, the system costs. For the underlying methodology, see \citet{Goeke2022}. 
    \item  \textbf{Space and process heating}. Nine technologies can meet the final demand for space heating, 28 for process heating, and 19 for transport services. We impose a must-run constraint on the technologies providing space heat, process heat, or transport services, enforcing a generation profile proportional to the demand profile \citep{Goeke2023}. This approach reflects that the demand of an individual consumer determines the operation of end-user technologies. Since they are not part of an extended network, these technologies do not have the flexibility to operate as base- or peak-load plants. Furthermore, there are 22 technologies for district heat, which serves as an intermediate carrier to meet the final demand for space or process heat below 100°C. A technology that converts district heat to space or process heat represents heating networks. Like PV and wind, the technology has three categories representing areas with low, average, and high network expansion costs, parametrized based on \citet{Fallahnejad2024}. 
    \item  \textbf{Biomass and synthetic fuels}. A total of 27 technologies cover the conversion of biomass and synthetic fuels. In addition to hydrogen generation from electricity, different pathways allow for producing methanol, synthetic gas and oil, and high-value chemicals, using biomass or direct air capture as a carbon source. The model includes three methanol production pathways: hydrogen with air-captured carbon, hydrogen with biogas, or solid biomass. Synthetic gas has three production pathways: raw biogas, raw biogas and hydrogen, or solid biomass. Production of synthetic oil is possible from solid biomass, hydrogen with air-captured carbon, or methanol. For each route, the produced crude oil must be refined at a 7\% energy loss to account for further processing into kerosene, naphtha, diesel, and gasoline \citep{concawe2012}. The production of high-value chemicals requires refined oil or methanol. For parametrization, we rely on input data from the PyPSA technology database for conversion of methanol to high-value chemicals, crude oil to high-value chemicals, and methanol to crude oil, since Danish Energy Agency data does not cover these key technologies \citep{DEA,PyPSAdata}.
\end{itemize}
Cost and performance assumptions for technologies build on the data published by the Danish Energy Agency \citep{DEA} for 2040. To reduce the number of investment variables in the subsequently described stochastic optimization, we first solve a deterministic model for each of the 35 climatic years in the underlying sample, described in the third section. Then, we check the model results for technologies that see no investment in any of the 35 years and exclude them from further analysis. This method reduces the number of considered conversion technologies from 101 to 49.

Beyond conversion, the model includes nine storage technologies, again building on data from the Danish Energy Agency \citep{DEA}. Lithium batteries for electricity storage and water tanks for district heat are exclusively short-term storage and can only shift energy within a month. Pit thermal storage for district heat, and storage tanks for hydrogen, can serve as long-term seasonal storage. Finally, hydro reservoirs, pumped hydro storage, caverns for hydrogen storage, and large-scale systems for synthetic or fossil gas and oil are available for multi-year storage. For details on the storage modeling, see the following subsection.

In addition to conversion and storage, the model includes five technologies to exchange energy between regions. HVAC (high-voltage alternating current) or HVDC (high-voltage direct current) lines enable electricity transport between regions. Due to its long technical lifetime, today's grid capacity is available without additional investments, but additional expansion incurs investment costs according to the capacity-cost curves published in \citet{entsoe2020}. Furthermore, the model can invest in a newly constructed hydrogen network and transport biomass and oil between regions at variable costs.

Finally, the model can leave the final demand unmet instead of using the various technologies to cover it. Unmet demand is subject to opportunity costs of 13'000 €/MWh, an estimate for the economic costs of unmet energy based on \citet{Ropke2013}. Since the costs of unmet demand are probability weighted in the stochastic optimization, the option of unmet demand prevents an overengineered system hedged against a virtually impossible event.

The model varies temporal and spatial resolution by energy carrier to reduce the model size but still achieve a high level of detail where it is most sensible \citep{Goeke2020b}. We apply an hourly resolution for electricity, representing all 8760 hours of the year. For space and district heat, the model applies a four-hour resolution capturing the thermal inertia of buildings and heating networks; process heat uses the same resolution to reflect demand-side flexibility. Demand for road transport has a daily resolution, but due to the hourly resolution for electricity, an hourly profile restricts charging and reflects when BEVs connect to the grid. Within these constraints, BEVs are flexible in covering their daily demand. The model uses a daily resolution for hydrogen and synthetic or fossil gas to reflect the carriers' inertia and the pipelines' storage capabilities. Finally, we deploy a monthly resolution for synthetic or fossil gas due to its comparatively low storage costs. For an in-depth discussion of deployed resolutions, see \citet{Goeke2023}.

\begin{figure}[h]
 \centering
  \includegraphics[scale=0.65]{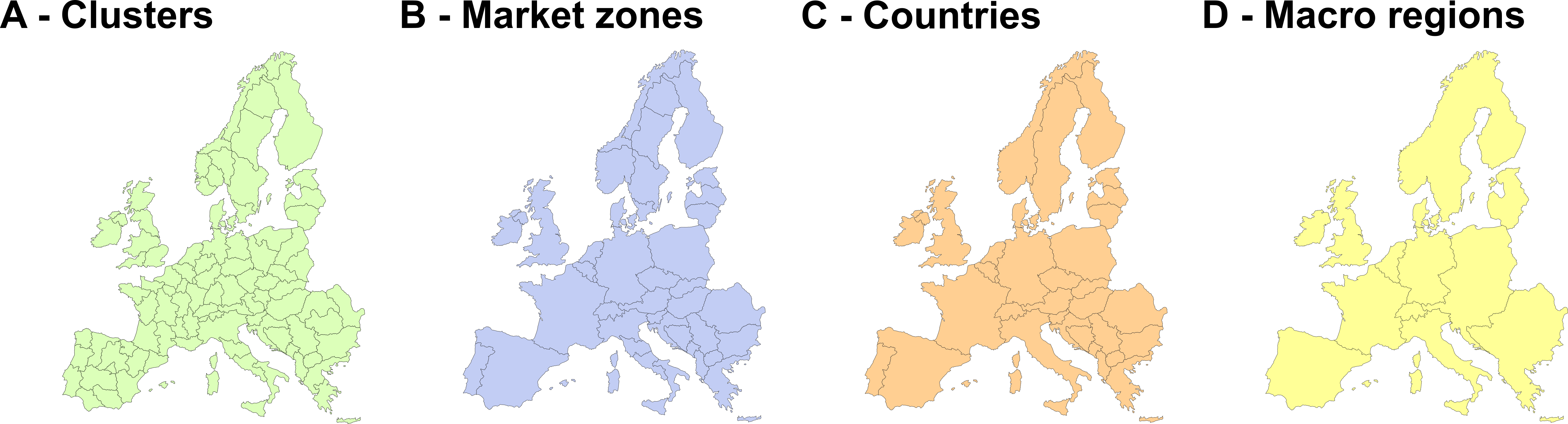}
 \caption[LoF entry]{The maps show the four spatial resolutions in the model. Investment and operation of onshore wind use the 96 clusters; offshore wind, PV, and hydro use the 46 market zones; liquid and gaseous fuels use the 11 macro-regions; and everything else uses the 33 countries.}
 \label{fig:map}
\end{figure}

Spatially, the model uses the four different resolutions shown in Fig. \ref{fig:map}. The finest resolution corresponds to the 96 clusters in (A). The model uses this resolution for capacity and dispatch decisions of onshore wind, the technology with the greatest spatial fluctuations. The next resolution in (B) corresponds to the 46 market zones of the European power market and is used for offshore wind, PV, and hydro plants. The country-level resolution in (C) balances supply and demand for electricity, heat, biomass, carbon as a commodity, and transport services. Accordingly, the model neglects bottlenecks and transport restrictions for these carriers within each country. Finally, the coarsest resolution is the 11 macro regions in (D) used for the energy balance of all fuels, including hydrogen, methanol, gas, and oil. This assumption implies that the model decides separately on the investment and operation of long-term storage for these energy carriers in each macro-region.

The choice of spatial resolutions balances technical accuracy, data limitations, and computational tractability. High resolutions are generally preferable for technical accuracy, but can significantly reduce data quality and increase model size. For instance, if a model can not achieve extreme spatial detail, the representation of exchange infrastructures is limited to aggregated or abstracted power grids and hydrogen networks, which are difficult to parametrize accurately. Our application prioritizes high spatial resolutions for renewable generation since spatial fluctuations are substantial and corresponding data is available. We apply a lower resolution for gaseous and liquid fuels since spatial dependence is smaller, data is limited, and higher resolutions are computationally expensive due to the stochastic approach described in the subsequent sections.

\subsubsection*{Supply contracts for fuel imports}

For imports of fuels from outside Europe, the model assumes fixed supply contracts with limited flexibility. In previous research, imports were entirely flexible and only subject to variable costs. However, in reality, energy markets run on long-term obligations, and, as the gas crisis demonstrated, market liquidity is insufficient to rely on imports in the short term.

Against this background, we introduce a method for fuel imports based on current long-term supply contracts for gas and other fuels. A take-or-pay clause in these contracts obligates the buyer to pay for quantities not taken, typically between 70\% and 95\% of the agreed price. Furthermore, volume tolerances allow buyers to reduce quantities, typically not by more than 10\% \citep{Ason2022,Creti2004}. Accordingly, our representation of supply contracts in Eq. \ref{eq:1a_1} to \ref{eq:1a_3} has an inflexible baseline plus an option for flexible imports. The fixed parameter $\theta$ provides the maximum flexible imports relative to the baseline, corresponding to the volume tolerance. The parameters $\nu_{base}$ and $\nu_{flex}$ are the marginal costs of baseline and flexible imports, respectively.
\begin{subequations}
\begin{alignat}{2}
y_{base} \leq  & \, x_{fuel} \cdot 8760 \label{eq:1a_1} \\
y_{flex} \leq  & \, x_{fuel}  \cdot 8760 \cdot \theta  \label{eq:1a_2} \\
C_{fuel} = & \,  x_{fuel} \cdot 8760 \cdot \nu_{base} + \nu_{flex} \cdot y_{flex}  \label{eq:1a_3}
\end{alignat}
\end{subequations}
Based on the cited literature values for supply contracts, $\theta$ is 10\% and $\nu_{flex}$ is 20\% higher than $\nu_{base}$ in the reference case. In the case of high import flexibility, $\theta$ is 30\% and $\nu_{flex}$ is 5\% higher than $\nu_{base}$.

\begin{figure}[h]
 \centering
  \includegraphics[scale=0.65]{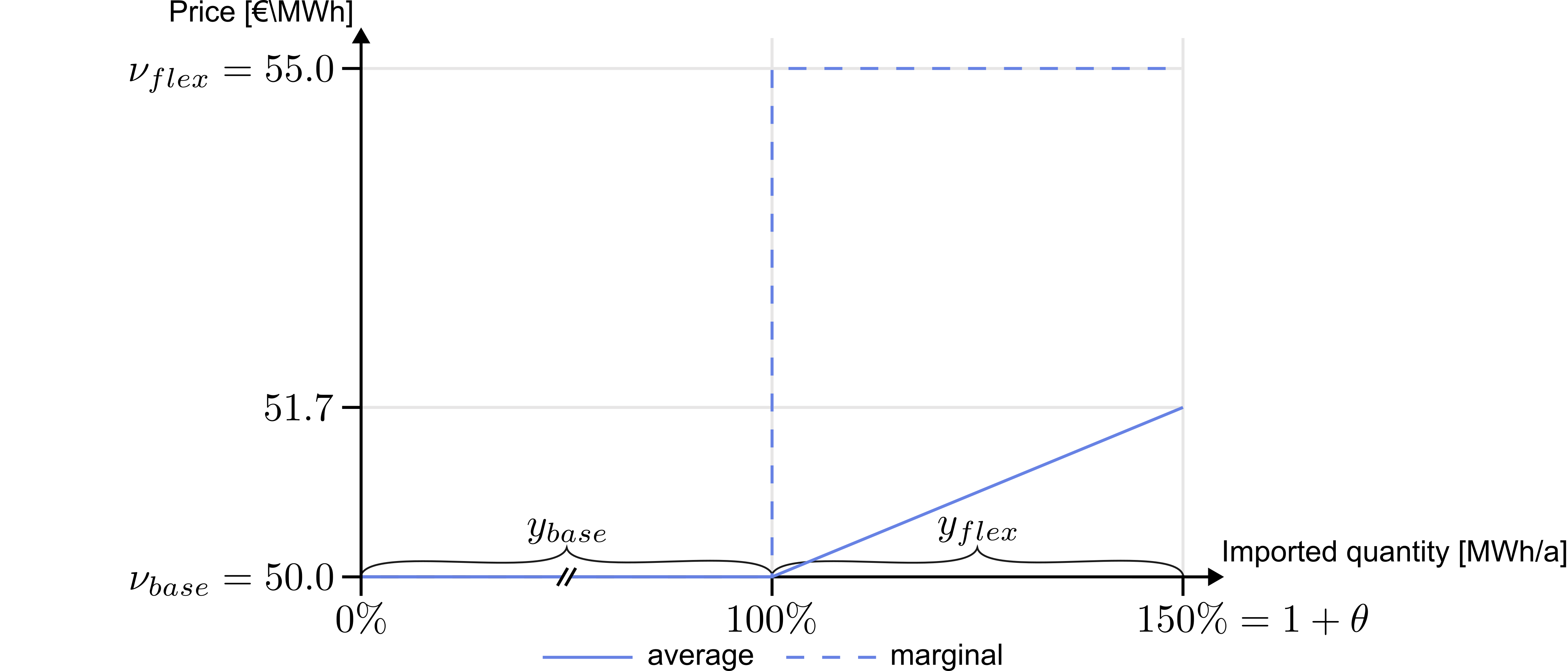}
 \caption[LoF entry]{For supply contracts that model fuel imports from outside Europe, the average and marginal import price depend on the quantities imported. For the inflexible baseline up to 100\%, the average and marginal prices correspond to the baseline costs $\nu_{base}$, namely 50.0 €/MWh in the example. For flexible imports exceeding the baseline, marginal costs increase to $\nu_{flex}$, or 55.0 €/MWh in the example. As a result, the average import price increases to 51.7 €/MWh when fully exploiting the options of flexible imports whose share is limited by $\theta$, corresponding to 50\% in the example.}
 \label{fig:impCon}
\end{figure}

The inflexible baseline of imports is a first-stage decision, analogous to the capacity expansion of technologies. Accordingly, the capacity variable $x_{fuel}$ in Eq. \ref{eq:1a_1} determines the baseline import quantity $y_{base}$. In the second stage, the model can add flexible imports $y_{flex}$ up to the prescribed share, enforced in Eq. \ref{eq:1a_2}. The total import costs $C_{fuel}$ in Eq. \ref{eq:1a_3} consist of a fixed base term that depends on the first-stage contract capacity and a flexible term related to the additional second-stage imports. Fig. \ref{fig:impCon} shows the resulting dependence of average and marginal prices on import quantities.

In a deterministic single-year model, modeling imports as flexible or with the proposed supply contracts makes no difference. The yearly demand will strictly correspond to the inflexible baseline to minimize costs. However, in the stochastic model introduced in the following section, varying imports can be beneficial to adapt to a surplus or deficit from renewables, and a trade-off between costs and flexibility arises.

\subsection*{Stochastic model for robust planning under climate uncertainty and limited foresight} 

This section introduces the refinements to the standard planning model to consider climate uncertainty and multi-year storage. The first section describes the modeling challenges and the current status quo. The second section introduces the concept of climate as a stochastic process for energy modeling, and the third section details the implementation of this concept in the planning model.

\subsubsection*{Challenges and status-quo}

Climate has three stochastic characteristics that are a challenge for energy modelling:
\begin{enumerate}
 	\item  \textbf{Large stochastic population}. The range of physically conceivable climate years is infinite. Capturing this range sufficiently is difficult, as datasets typically only span up to 60 climate years \citep{Grochowicz2023,Grochowicz2024,Gotske2024}. In addition, computational limitations restrict the sample size a model can implement. At the same time, a poor representation of the overall population is at risk of missing extreme climate years that threaten reliability.
	\item \textbf{Continuity}. Climate evolves continuously, and different climate years follow one another. Analysing single years in isolation misses the risk of multi-year droughts and the potential for multi-year storage. 
	\item \textbf{Limited foresight}. Reliable weather forecasts are only possible a few weeks into the future. Climate conditions months or years into the future are unpredictable, and as a result, long-term storage must operate under uncertainty. 
\end{enumerate}

To represent climate impacts, the planning model introduced in the previous section requires climate-sensitive input data on capacity factors for wind and PV, inflows to hydro reservoirs, and the final demand for heat or electricity. These inputs are cross- and autocorrelated since they all result from the physical climate system. Preserving these correlations is crucial, but an analytical representation of atmospheric physics is too complex, so models take a data-driven approach, implementing time series from historical re-analysis or future simulations within the operational constraints in Eq. \ref{eq:1c}.

\begin{figure}[h]
 \centering
  \includegraphics[scale=0.65]{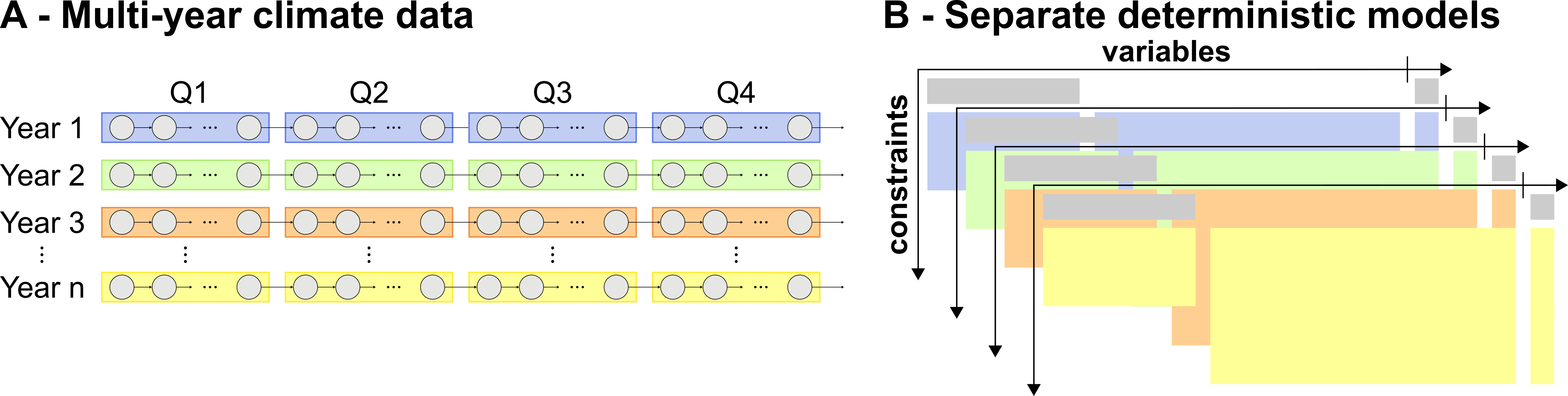}
 \caption[LoF entry]{The status quo of analyzing long-term climate impacts on energy systems is to (A) compile a multi-year sample of climate data and (B) run a deterministic planning model separately for each year. Then, heuristics search for a reliable setup among the results.}
 \label{fig:statusQuo}
\end{figure}

The state-of-the-art for considering long-term climate impacts is using a sample with multiple years of time-series data, symbolized by part (A) of Fig. \ref{fig:statusQuo}, and running separate perfect foresight models on parts of the data, symbolized by part (B) \citep{Grochowicz2023,Grochowicz2024,Gotske2024}. As a result, the state-of-the-art does not capture any of the three stochastic characteristics of climate listed above.

Recent studies extend the deterministic model to a stochastic model that includes multiple climate years \citep{Goeke2024,DeMarco2025}, described by Eqs. \ref{eq:2a} to \ref{eq:2d}. In the stochastic model, first-stage decisions determine capacities for multiple climate years of operation in the second stage. Correspondingly, the first-stage variables for investment remain the same, and the second-stage variables for operation, $y_s$, depend on the scenario or climate year, $s$. The objective function Eq. \ref{eq:2a} minimizes expected costs over all scenarios, factoring in their probability $\rho_{s}$. Accordingly, the costs of unmet demand are also probability weighted, preventing disproportionate costs to secure the system against unlikely extremes. Operational constraints in Eq. \ref{eq:2c} are enforced separately for each scenario.

\begin{subequations}
\begin{alignat}{2}
\min_{x,y} \; \; & c^\top x + \sum_{\mathclap{\,s \in S}} \; \rho_{s} && \cdot \, d_{s}^\top \, y_{s} \label{eq:2a}  \\
s.t. \; \; & H x && \leq  a \label{eq:2b} \\  
& I_{s} \, x + J_{s} \, y_{s} && \leq b_{s} \; \; \forall \, s \in S  \label{eq:2c} \\  
& x \in \mathbb{R}^{n}_+,\, y_{s} \in \mathbb{R}^{n}_+ \label{eq:2d}  
\end{alignat}
\end{subequations}

Although the stochastic model can consider multiple climate years, computational limitations still restrict the sample to a size too small to describe the entire range of climate conditions. In addition, the stochastic model neglects the continuity of climate, since it analyzes different climate years, or scenarios, in isolation---not in succession. Third, the stochastic model does not limit foresight, since within each scenario, there is perfect foresight of climate conditions for the entire year.

\subsubsection*{Climate as a stochastic process in energy modeling}

Next, we describe our methodology for overcoming the status quo and capturing the three stochastic climate characteristics. As a basis, this section describes climate as a stochastic process in an energy modeling context. The following sections will apply this concept to refine the stochastic model further.

First, we assume that the climate is uncorrelated or independent over sufficiently long periods. For instance, the weather in August does not indicate how cold the winter will be. For introducing the method, we define a sufficiently long period as three months or one quarter. In the next section, we revisit the assumption and discuss it in light of empirical evidence. Second, we assume that less than one quarter into the future, foresight is perfect and discrete rather than rolling. So, at every timestep of a three-month period starting on January 1st, there is perfect information until March 31st, but no information beyond March 31st. Then, on April 1st, perfect information for the following three months is revealed. Such discrete foresight is implausible, but a significant improvement over perfect foresight. Additionally, it primarily biases short-term operations, which is tolerable since the focus of this work is long-term storage.

\begin{figure}[h]
 \centering
  \includegraphics[scale=0.65]{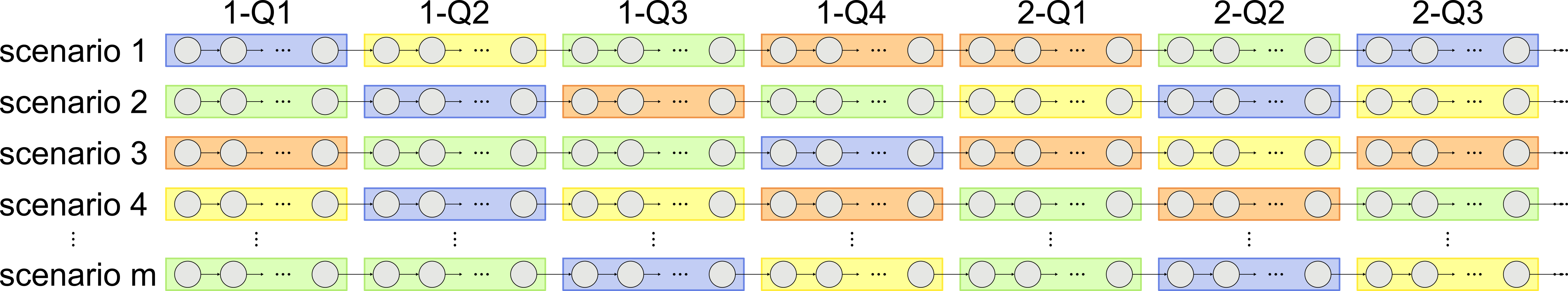}
 \caption[LoF entry]{Bootstrapping can extend the existing sample and provide a multi-year time series if we describe weather as a continuous stochastic process with little dependence over long periods. To illustrate the idea, the scenarios in the example originate from drawing random quarters from part (A) of Fig. \ref{fig:statusQuo}, preserving the order of seasons.}
 \label{fig:stochasticNature}
\end{figure}

Applying the concept of climate uncertainty as a stochastic process, we use bootstrapping to generate new samples. To this end, we draw random quarters from the data in part (A) of Fig. \ref{fig:statusQuo}, preserving the order of seasons. The approach generates any number of samples spanning an arbitrary number of years, resulting in the extended sample illustrated in Fig. \ref{fig:stochasticNature}. As a result, the extended sample resolves the challenges of having an exhaustive sample and considering the continuity of climate conditions. The approach comes at the cost of imposing some discontinuity, for example, transitioning from a cold January 31st to a warm February 1st. Again, these discontinuities primarily bias short-term operations, which are not the focus of this work.

\subsubsection*{Scalable implementation}

In practice, however, simply running the stochastic model on an extended sample with hundreds of decade-long scenarios is impossible due to computational limitations. Therefore, we introduce a scalable combinatorial implementation of the stochastic process to consider climate uncertainty in the planning problem. This approach splits the scenarios by periods, in the current example, quarters, adding the index $p$ for the period to all second-stage variables and constraints, which results in Eqs. \ref{eq:3a} to \ref{eq:3d}. 

\begin{subequations}
\begin{alignat}{2}
\min_{x,y} \; \; & c^\top x + \sum_{\mathclap{p \in P,\,s \in S}} \; \rho_{p,s}  \cdot \, d_{p,s}^\top && \, y_{p,s}  \label{eq:3a} \\
s.t. \; \; & H x && \leq  a \label{eq:3b} \\  
& I_{p,s} \, x + J_{p,s} \, y_{p,s} && \leq b_{p,s} \; \; \forall \, p \in P, s \in S   \label{eq:3c} \\  
& x \in \mathbb{R}^{n}_+,\, y_{p,s} \in \mathbb{R}^{n}_+ \label{eq:3d}
\end{alignat}
\end{subequations}

The fundamental idea of splitting into period-scenario combinations is having a manageable number of scenarios while ensuring the model considers all combinations of these scenarios to cover a broad range of climate conditions. We can quantify the number of period-scenario combinations for the entire year $n_{tot}$ as the product of the scenarios in each period $n_p$: 

\begin{alignat}{3}
n_{tot} = & \prod_{p \in P} n_p \label{eq:growth}
\end{alignat}

\begin{figure}[h]
 \centering
  \includegraphics[scale=0.65]{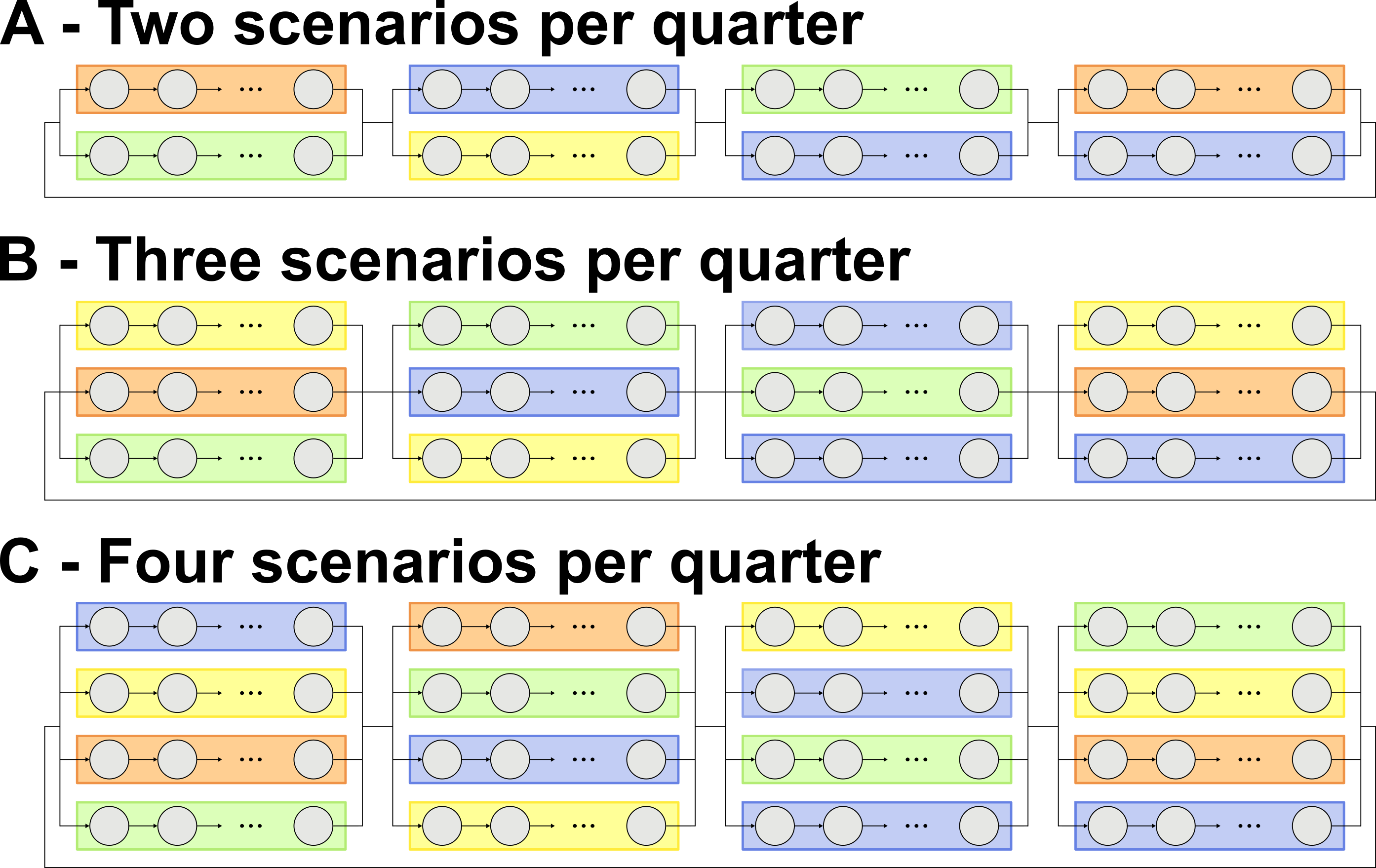}
 \caption[LoF entry]{Our idea is to leverage exponential growth and model only a manageable number of scenarios per period, but guarantee reliability for all combinations of scenarios. In part (A), two scenarios per quarter result in 16 combinations for a single year, or 16 paths to move through the graph from left to right; in part (B), three scenarios per quarter already result in 81 combinations, and in part (C), four scenarios result in 256 combinations.}
 \label{fig:combination}
\end{figure}

Correspondingly, a model with two scenarios per quarter, shown in part (A) of Fig. \ref{fig:combination}, leads to 16 combinations for the entire year. Suppose the orange year has a cold winter in quarters four and one, and the blue year has a summer with little excess supply in quarters two and three. The four over two, viz. 16, combinations will include blue in winter and orange in summer---the worst case, since discharging of storage during winter is high, but charging during summer is low. Thanks to the probability parameter $\rho_{p,s}$, the combinatorial model still rates the likelihood of each scenario. The following section on the clustering algorithm describes our approach for computing the probability. 

According to Eq. \ref{eq:growth}, the number of climate scenarios for a single year grows exponentially. For three scenarios per quarter in part (B) of Fig. \ref{fig:combination}, the combinations add up to 81 for the entire year; for four scenarios in part (C), the combinations add up to 256. Shortening the periods has an even greater impact. Moving from quarterly to monthly periods with four scenarios per period has combinations surging from 256 to 16'777'216 at the same model size.

\subsubsection*{Robust formulation for long-term storage} \label{modelImpl}

The description of our methodology in the previous section explains how the refined model can consider an extensive climate sample, but not how to achieve continuity and limited foresight. In practice, implementing continuity and limited foresight comes down to refining the model for long-term storage, the only model element that connects the different periods.

In this section, we introduce a robust formulation that splits long-term storage into a seasonal and a multi-year component. The formulation explicitly tracks the level of the seasonal component, but for the multi-year component, the combinatorial approach introduced in the previous section prevents explicit tracking. Therefore, we develop a scalable and robust formulation for the multi-year component that ensures discharged and charged energy are consistent in the long run and the storage size is sufficient to balance an extreme surplus or deficit in the short run.

All storage modeling builds on the standard equations in Eqs. \ref{eq:5a} to \ref{eq:5d} (for clarity, the equations omit efficiency parameters). The storage level ${lvl}_{t}$ in timestep $t$ equals the level in the previous timestep plus the delta of energy charged and discharged to the energy balance, ${in}^{bal}_{t}$ and ${out}^{bal}_{t}$, respectively. Furthermore, charging, discharging, and storage levels are subject to capacity constraints linking them to respective capacity variables.

\begin{subequations}
\begin{alignat}{4}
{lvl}_{t}  = & \; {lvl}_{t-1}  &&  + \, {in}_{t} - {out}_{t} && \; \; \forall \, t \in T  \label{eq:5a} \\ 
0 \, \leq & \, \, \, \, \,   {in}_{t} && \leq x^{in} && \; \; \forall \, t \in T \label{eq:5b}  \\ 
0 \, \leq & \, \, \, \, {out}_{t}   && \leq x^{out} && \; \; \forall \, t \in T  \label{eq:5c}\\
0 \,  \leq & \, \, \, \, \,  {lvl}_{t}  && \leq x^{size} && \; \; \forall \, t \in T \label{eq:5d}
\end{alignat}
\end{subequations}
In the model, short-term storage, like batteries, use precisely these equations plus a cyclic constraint that enforces the same level at the start and end of each period $p$.

Long-term storage uses the refined storage equations in Eqs. \ref{eq:6a} to \ref{eq:6c} instead of Eqs. \ref{eq:5a} to \ref{eq:5d}. Storage levels are still computed from the delta to the previous timestep $t$, but this delta has two origins now. First, as in the standard approach, the charging and discharging to the overall energy balance, ${in}^{bal}$ and ${out}^{bal}$, which is still subject to capacity restrictions. Second, as new terms, charging and discharging to a multi-year storage, ${in}^{mul}$ and ${out}^{mul}$.
\begin{subequations}
\begin{alignat}{3}
{lvl}_{t,s}  = & \, {lvl}_{t-1,s} \, + \, {in}^{bal}_{t,s} - {out}^{bal}_{t,s} + {in}^{mul}_{t,s} - {out}^{mul}_{t,s} && \; \; \forall \, t \in T, s \in S  \label{eq:6a} \\ 
x^{in} \, \, \geq & \, {in}^{bal}_{t,s} && \; \; \forall \, t \in T, s \in S \label{eq:6b}  \\ 
x^{out} \geq & \, {out}^{bal}_{t,s} && \; \; \forall \, t \in T, s \in S  \label{eq:6c}
\end{alignat}
\end{subequations}
Adding the terms for a multi-year storage, we can split long-term storage into a component for seasonal storage and a component for multi-year storage. Unlike the distinction between short-term and long-term storage, this split into seasonal and multi-year is not physical but for accounting in the model, and not exogenous but endogenous to the model.

\begin{figure}[h]
 \centering
  \includegraphics[scale=0.65]{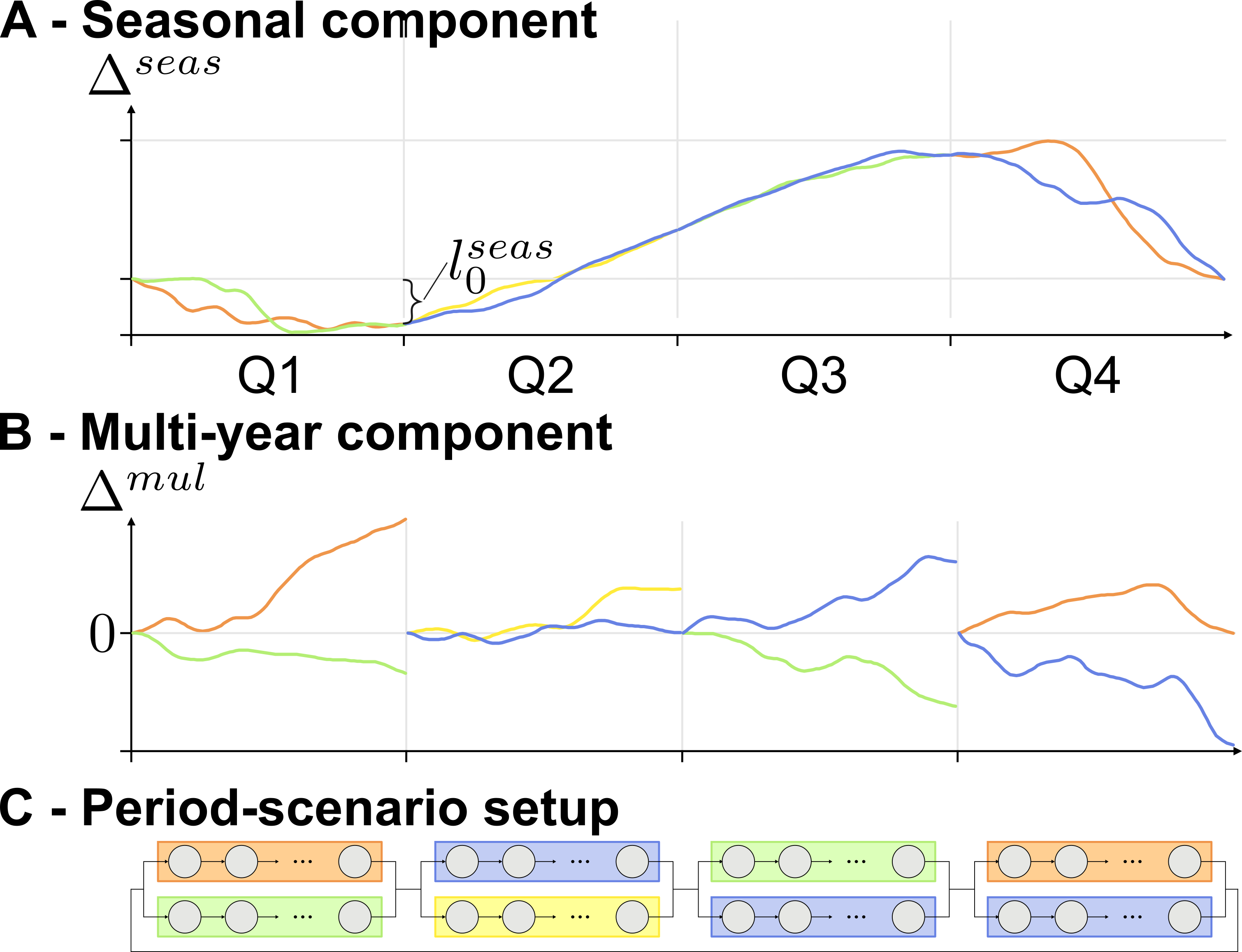}
    \caption[LoF entry]{Long-term storage consists of two components, as shown in the example period-scenario setup in part (C). For the seasonal component in part (A), the total delta is the same for each scenario at the same quarter, creating a uniform seasonal profile. For the multi-year component in part (B), the delta can differ by scenario, enabling multi-year storage.}
 \label{fig:storage}
\end{figure}

The seasonal component has the same delta for all scenarios of the same period, viz., it is the same for every climate year, as stated in Eq. \ref{eq:7a}. Indirectly, the constraint enforces equal storage levels at the start or end of each period $p$ spanning several timesteps $t$, creating a seasonal profile as illustrated in part (A) of Fig. \ref{fig:storage}.
\begin{alignat}{4}
\Delta_{p}^{seas} = & && {lvl}_{p,s} - {lvl}_{p-1,s} && \; \; \forall \, p \in P, s \in S \label{eq:7a} 
\end{alignat}
Since levels are the same at the end of a period, the operation does not anticipate which scenario will materialize in the coming period. Furthermore, levels will always be sufficient to manage the worst-case. For instance, if the blue scenario for the fourth quarter in Fig. \ref{fig:storage} has high demand and requires substantial energy from long-term storage, the third quarter will charge the storage accordingly, even if the orange scenario with less demand materializes instead. Vice versa, if the green scenario in the third quarter has little excess supply, the constraint still ensures it has a sufficient surplus to charge the storage, even if the blue scenario with a larger surplus materializes instead. Therefore, the seasonal component alone is sufficient for reliable planning with limited instead of perfect foresight. 

However, with the seasonal component alone, reliability comes at the expense of substantial excess energy. For instance, if the blue scenario with less demand materializes for the fourth quarter, Eq. \ref{eq:7a} still forces storage discharge, although there is no corresponding demand. Vice versa, if the blue scenario with high surplus materializes for the third quarter, Eq. \ref{eq:7a} prevents additional charging and forces curtailment. In conclusion, the seasonal component prohibits setting aside reserves in years with a climate-driven surplus and drawing on these reserves in years with a deficit.

To address this limitation, the multi-year storage component complements the model of long-term storage. The delta of the multi-year component in Eq. \ref{eq:7b} is the total energy charged and discharged to the multi-year storage corrected by the parameter $\eta$ to account for long-term losses. Unlike the seasonal component, the delta of the multi-year storage may vary across scenarios for the same period. As a result, the multi-year component enables upward or downward deviations relative to the uniform seasonal profile, as shown in part (B) of Fig. \ref{fig:storage}.
\begin{alignat}{4}
\Delta_{p,s}^{mul} \, \,  = & \, \sum_{\mathclap{t \in P,\,s \in S}} \; && \eta \cdot {out}^{mul}_{t,s} -  {in}^{mul}_{t,s} &&  \; \; \forall \, p \in P, s \in S \label{eq:7b}
\end{alignat}

To control the multi-year component, we cannot explicitly track its storage level. Unlike the seasonal component, the level depends on the realization of uncertainty in the preceding periods and years. As a result, tracking its level requires an extensive scenario tree with all combinations. However, such a tree contradicts our combinatorial idea of implicitly considering all these combinations yet not modeling them to keep the model tractable. Therefore, we pursue a robust approach to model the multi-year storage, ensuring its sizing and level are sufficient even in extreme cases.

First, Eq. \ref{eq:8} ensures that the multi-year storage does not discharge more energy than it charges in the long run. The constraint sets a lower limit on its expected level defined by the sum of the probability-weighted deltas across all period-scenario combinations. The lower limit is the variable for the starting level of the multi-year storage $l_0$ times the parameter $\alpha$.
\begin{alignat}{3}
\alpha \cdot l_0  \leq & \, \sum_{\mathclap{p \in P,\,s \in S}} \; \rho_{p,s} \cdot \Delta_{p,s}^{mul} && \label{eq:8} 
\end{alignat}The parameter $\alpha$ serves as a safety margin, and it is the inverse of the average time in years it takes for a fully depleted storage to return to the starting level.

However, even if the storage level is positive in the long run, a prolonged series of adverse weather can still deplete it. Therefore, additional constraints ensure that the starting level and storage size are sufficient to balance an extreme deficit or surplus in the short run.

\begin{figure}[h]
 \centering
  \includegraphics[scale=0.65]{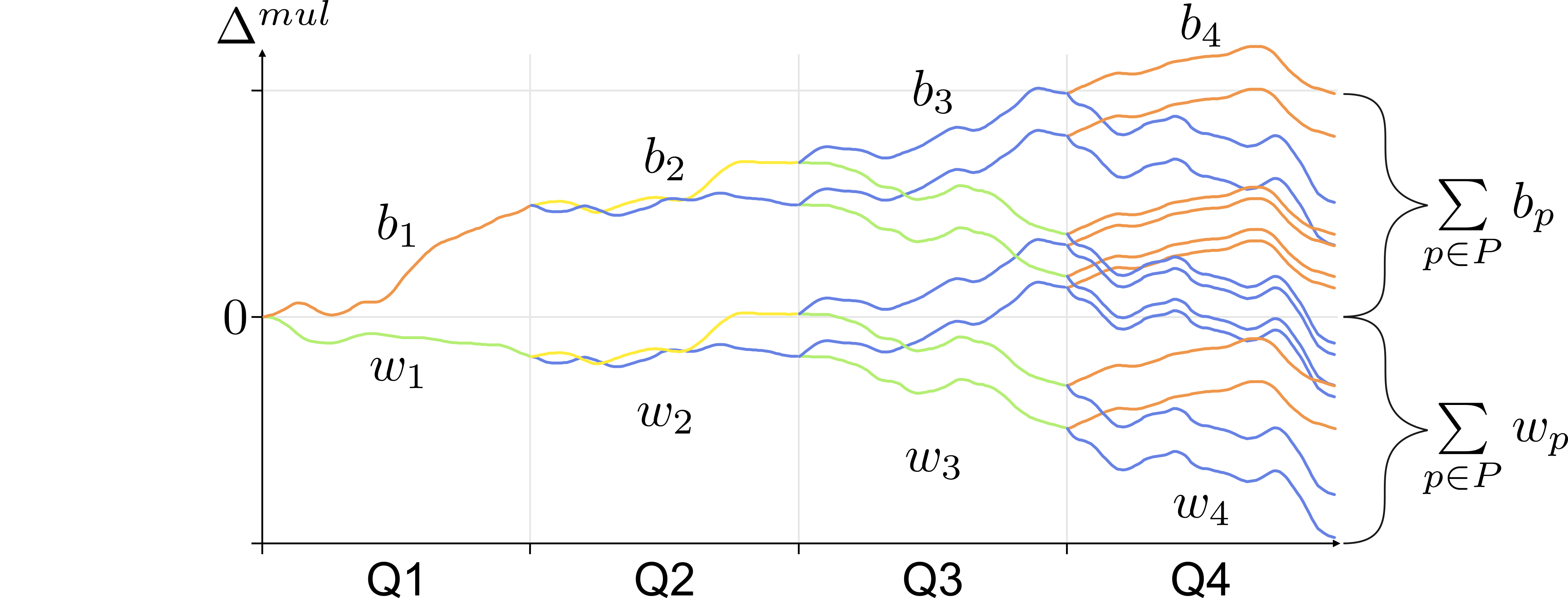}
 \caption[LoF entry]{The delta for the multi-year component of long-term storage varies by scenario. As a result, all possible values for the delta of an entire year result from the scenario tree in the figure. The scenarios with the smallest delta constitute the worst-case path at the bottom; the ones with the largest delta constitute the best-case path at the top.}
 \label{fig:worstCase}
\end{figure}

As a first step, first, Eqs.\ref{eq:9a} and \ref{eq:9b} identify the worst-case delta $w_{p}$ for each period, forcing $w_{p}$ to the smallest delta value across all scenarios, the lower line for each quarter in part (B) of Fig. \ref{fig:storage}. The worst-case corresponds to the bottom path in the scenario tree in Fig. \ref{fig:worstCase}. Analogously, to prevent spillage, Eqs.\ref{eq:9c} and \ref{eq:9d} define the best-case with the largest delta for each period, the upper line for each quarter in part (B) of Fig. \ref{fig:storage}, resulting in the top path in the scenario tree.
\begin{subequations}
\begin{alignat}{3}
w_{p}  \leq  & \; \Delta_{p,s}^{mul}  && \; \; \forall \, p \in P, s \in S \label{eq:9a} \\
w_{p}  \leq  & \; 0  && \; \; \forall \, p \in P, s \in S \label{eq:9b} \\
b_{p}  \geq  & \; \Delta_{p,s}^{mul}  && \; \; \forall \, p \in P, s \in S \label{eq:9c} \\
b_{p}  \geq  & \; 0 && \; \; \forall \, p \in P, s \in S \label{eq:9d}
\end{alignat}
\end{subequations}

On this basis, Eqs. \ref{eq:10} to \ref{eq:10a0} assure a sufficient starting level $l_0$ for the multi-year storage to prevent depletion. Eq. \ref{eq:10} defines the starting level as the sum of the starting level of the multi-year component $l^{mul}_0$ and the seasonal component $l^{seas}_0$.

Eq. \ref{eq:10a} ensures the starting level of the multi-year component $l^{mul}_0$ is sufficient to prevent depletion if the worst-case year repeats $\beta$-times. To achieve this, it sets a starting level greater than the aggregated worst-case delta $w_{p}$ for an entire year times the parameter $\beta$ that serves as a safety margin.

\begin{figure}[h]
 \centering
  \includegraphics[scale=0.65]{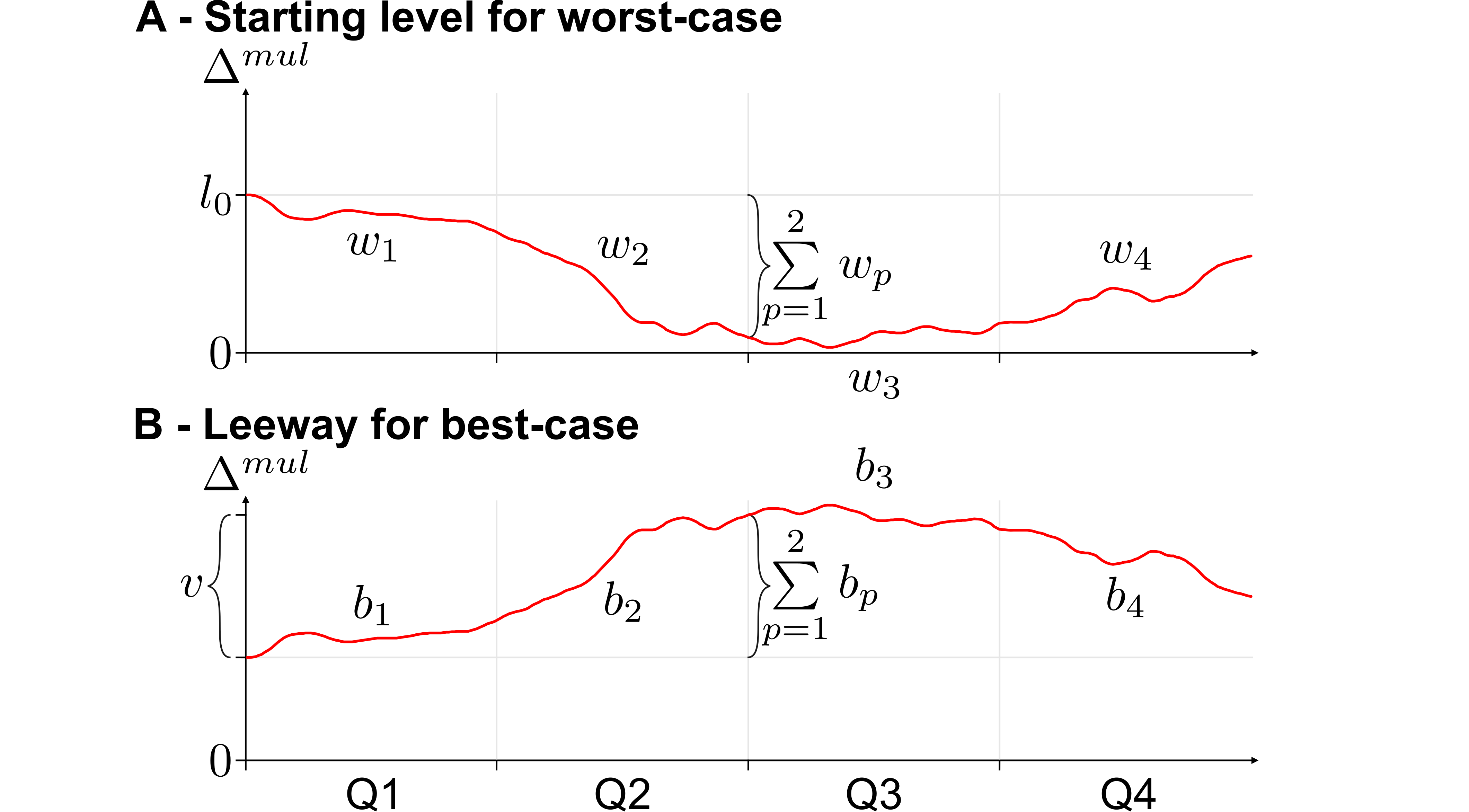}
 \caption[LoF entry]{The multi-year storage's starting level and energy capacity must consider the worst-and best-case path over one year. First, in part (A), within the year, the worst-case path can be below its final value at the end of the year. In this case, the starting level must increase. Second, in part (B), within the year, the best-case path can be above the starting level. Here, the storage needs sufficient leeway to store the additional energy.}
 \label{fig:overUnder}
\end{figure}

However, this constraint is necessary but not sufficient. Depletion can still occur if, within the worst-case year, the delta $w_{p}$ changes from negative to positive, and, as a result, the aggregated delta is temporarily below the yearly value, as shown in part(A) of  Fig. \ref{fig:overUnder}. To prevent this, Eq. \ref{eq:10b} additionally ensures that the starting level is sufficient to cover the first period, the first plus the second period, the first plus the second plus the third period, and so on, by summing the respective delta $w_{p}$. Analogously to Eq. \ref{eq:10b}, Eq. \ref{eq:10a0} prevents temporary depletion for the seasonal component.

\begin{subequations}
\begin{alignat}{4}
l_0 \;  \;  \: \:     = & \; \; \; \; l^{seas}_0  \; + && \; \; l^{mul}_0 && \label{eq:10} \\
l^{mul}_0 \: \, \geq &  - \beta \; \; \cdot  && \sum_{p \in P} w_{p} && \label{eq:10a} \\
l^{mul}_0 \;  \geq & -1 \; \; \cdot \;  && \sum_{p = 1}^{\hat{p}} w_{p}  && \; \; \forall \, \hat{p} \in P \label{eq:10b} \\
l^{seas}_0  \geq  & -1  \; \; \cdot && \sum_{p = 1}^{\hat{p}} \Delta_{p}^{seas} &&  \; \; \forall \, \hat{p} \in P \label{eq:10a0}
\end{alignat}
\end{subequations}

Analogously to the starting level, Eqs. \ref{eq:11a} and \ref{eq:11b} enforce a leeway $v$ of free storage capacity to prevent energy spillage. Eq. \ref{eq:11a} sets the leeway above the best-case delta for an entire year times the parameter $\gamma$, corresponding to the repetitions of the best-case without spillage. Eq. \ref{eq:11b} ensures sufficient leeway to prevent spillage within the year. The constraint applies when, within the year, the best-case path is above its final value at the end of the year, as shown in part (B) of Fig. \ref{fig:overUnder}.
\begin{subequations}
\begin{alignat}{3}
v  \geq \, \gamma  \cdot  & \sum_{p \in P} b_{p} && \label{eq:11a} \\
v \geq \; \; \; \; \, & \sum_{p = 1}^{\hat{p}} b_{p}  && \; \; \forall \, \hat{p} \in P \label{eq:11b}
\end{alignat}
\end{subequations} 
Note that Eqs. \ref{eq:10b} to \ref{eq:11b} are redundant if the worst-case delta $w_{p}$ is strictly negative and the best-case delta $b_{p}$ is strictly positive which is the case in the optimum. However, the constraints are still valuable in ensuring the consistency of sub-optimal solutions and restricting the solution space to speed up optimization.

The multi-year storage's required energy capacity $x^{mul}$, defined in Eq. \ref{eq:11a}, corresponds to the starting level plus the additional leeway $v$. 
\begin{alignat}{3}
x^{mul}   = & \; l^{mul}_0 + v && \label{eq:11c}
\end{alignat}

The energy capacity for the seasonal component $x^{seas}$ in Eq. \ref{eq:12a} is simply the upper limit for the explicit seasonal storage level. The sum of the seasonal and multi-year components gives the total energy capacity of long-term storage, as defined in Eq. \ref{eq:12b}.
\begin{subequations}
\begin{alignat}{3}
x^{seas}  \geq & \; {lvl}_{t,s} && \; \; \forall \, t \in T, s \in S \label{eq:12a}  \\
x^{size} \,   =  & \; x^{seas} + x^{mul} && \label{eq:12b} 
\end{alignat}
\end{subequations}
Being designed for the most extreme cases, utilization of the energy capacity is much smaller for the multi-year than for the seasonal component. In addition, using the multi-year storage is subject to additional losses, giving the cost-minimizing model an incentive to limit multi-year storage to deviations from the average seasonal profile.

The formulation for long-term storage addresses the remaining two challenges, continuity and limited foresight, since it accounts for multi-year storage or droughts and overcomes perfect foresight. However, the approach still has deficiencies. First, it relies on tuning the security margins $\alpha$ and $\beta$ to achieve the desired security of supply instead of enforcing it directly. Second, it cannot ensure that the storage levels fall below zero or exceed the energy capacity within a period, only at the start or end. For the computations in this paper, $\alpha$ is 1\% and $\beta$ is one year. Tests showed that on the one hand, these security margins are sufficient to prevent storage depletion reliably, but on the other hand, further reducing them did not have a noticeable impact on system costs.

\subsection*{Probabilistic selection of representative months}

The previous section described how the introduced model leverages bootstrapping and exponential growth to cover an extensive range of climate conditions while keeping the problem size manageable. The method requires subdividing the year into shorter stochastically independent periods and defining scenarios for each period. This section describes how we practically execute these two steps.

The underlying time series data is the Pan-European Climatic Database (PECD) version 2021.3 by ENTSO-E \citep{pecd}. The database builds on ERA5 re-analysis data and provides capacity factors for openspace PV, rooftop PV, onshore wind, offshore wind, and hydro generation from 1982 to 2015 \citep{Hans2020}. We extended the dataset with consistent heat demand and heat pump efficiency data by applying the methodology described in \citet{Ruhnau2019} and made the data publicly available \citep{pecdPlus}. For all other time series, we use climate-independent data \citep{plan4res}.

% discuss also plausibiliy from perpective that forecasts are possible for one month
We set the length of an independent period to one month, deviating from the examples in the previous section, which used a quarter. First, the empirical evidence supports this assumption since an autocorrelation analysis in \citet{Schmidt2025} on the PECD data shows that correlation is negligible after one month for PV and wind and small for hydro. Therefore, assuming one month of foresight is a conservative assumption. Second, shortening an independent period from a quarter to a month exponentially increases the combinations of climate conditions covered for the whole year, for instance, with two scenarios per period from 16 to 144.

The scenarios for each month are representative months from the 35-year dataset. The data for each month is high-dimensional, as it includes hourly data for capacity factors, hydro generation, heat demand, and heat-pump efficiency in each model region. At the same time, there are only 35 samples per month, so standard clustering methods, such as k-means, are not suitable. Instead, we adapt optimization-based clustering algorithms to our novel formulation \citep{Goeke2022b}.

Eqs. \ref{eq:16a} to \ref{eq:16d} provide the algorithm's underlying optimization problem. The algorithm decides on two binary variables: $u_{i,p}$ is one if month $p$ from year $i$ is one of the selected representative months, and zero if not. $v_{i,j,p}$ is one if month $p$ from year $i$ represents the same month from year $j$. For instance, if the algorithm selects January 2008 as a representative month, representing January from 2010 and 1994, $u_{\text{2008},\text{Jan}}$, $v_{\text{2008},\text{2010},\text{Jan}}$, and $v_{\text{2008},\text{1994},\text{Jan}}$ are one.

\begin{alignat*}{3}
 \textbf{Variables \;} & && \\
& u_{i,p}: && \; \text{one, if month $p$ from year $i$ is a selected as a representative month} \\
& v_{i,j,p}: && \; \text{one, if month $p$ from year $i$ represents month $p$ from year $j$} \\
 \textbf{Parameter} & && \\
& d_{i,j,p,c}: && \; \text{absolute difference regarding metric $c$ between month $p$ from year $i$} \\
& && \;  \text{and month $p$ from year $j$} \\
& n: && \; \text{total number of representative months} \\
&n^{\text{ext}}: && \; \text{number of representative months used for extreme months}
\end{alignat*}

\begin{subequations}
\begin{alignat}{6}
\min_{u,v}  &  && \sum_{\mathclap{ \substack{i \in I,\,j \in J,\,p \in P\\ c \in C}}}    && \; v_{i,j,p}  \, \cdot \, && d_{i,j,p,c} && \label{eq:16a} \\
&  && && \; v_{i,j,p} && \leq u_{i,p} && \forall \, i \in I, j \in J, p \in P \label{eq:16b} \\ % enfore relation between x and y
& && \sum_{i \in I}  && \; v_{i,j,p} && = 1 && \forall \, j \in J, p \in P \label{eq:16c} \\ % each month of each year must be represented
& \sum_{i \in I} && \sum_{p \in P} && \; u_{i,p}  && = n - n^{\text{ext}} && \label{eq:16d} \\ % enforce correct number of representative months
\noalign{\centering $v_{i,j,p} \in \{0,1\},\, u_{i,p} \in \{0,1\}$} \nonumber
\end{alignat}
\end{subequations}

The objective of the optimization in Eq. \ref{eq:16a} is to minimize the difference between the representative and represented months. The difference is computed from the parameter $d_{i,j,p,c}$, the absolute delta between month $p$ from year $i$ and year $j$ regarding a specific metric $c$. We consider eight metrics: the total supply from PV, onshore wind, offshore wind, and run-of-river, the inflows to hydro reservoirs and pumped storage, and the electricity demand of air-source and ground-source heat pumps. Since the time series data only provides relative supply and demand, computing total supply and demand requires installed capacities---resulting from the capacity expansion that requires representative months as an input. To overcome this circularity, we rely on an approximation and solve the model deterministically for each of the 35 years in the dataset, reducing the time series to 2'856 representative hours. We then use the resulting average capacities for clustering. With these capacities and the time series data, we compute the listed supply and demand metrics for each month and region, aggregate them by region, and compute their absolute delta between different months to obtain $d_{i,j,p,c}$.

The constraint in Eq. \ref{eq:16b} connects the two binary decision variables, ensuring that if a month represents another one according to $v_{i,j,p}$, also $u_{i,p}$ is one. Eq. \ref{eq:16c} enforces that each month from the full sample has a representative month. Eq. \ref{eq:16d} limits the total number of representative months selected by the algorithm to the total number of representative months $n$ minus the number of preselected extreme months $n^{\text{ext}}$.

The preselection is necessary since the optimization step aims to match the average of the representative and represented months and will level out the extremes as a result. An extreme month, for example, a January with high demand and low wind supply, will not be representative by definition and, therefore, never selected by the optimization. However, the month would be critical for reliable system planning. To identify extreme months, we compute the residual demand, meaning the renewable supply minus demand, with the same capacities as previously, and aggregate across all regions for the entire month. Then, we select $n^{\text{ext}}/2$ months with the highest and the  $n^{\text{ext}}/2$ months with lowest residual demand. The subsequent optimization-based clustering excludes these already selected extreme months.

From the results of the clustering algorithm, we can also derive probabilities of the representative periods, viz., months, used in the planning model. Assuming the original dataset is representative, the likelihood of a representative month corresponds to the share of months of the dataset it represents. 

Accordingly, we compute the probability of a representative month $\rho_{p,i}$, which corresponds to the probability of a period $\rho_{p,s}$ in the optimization model in Eqs. \ref{eq:3a} to \ref{eq:3d}. This probability described in  Eq. \ref{eq:17} is the ratio of months it represents in the numerator to the total number of months in the sample in the denominator.
\begin{equation}
\rho_{p,i} = \rho_{p,s} = \frac{\sum_{j} v_{i,j,p}}{\sum_{\hat{i},j} v_{\hat{i},j,p}} \label{eq:17} 
\end{equation}
The probabilities for each month of the year will always sum to one since each month is exactly represented once according to the constraint in Eq. \ref{eq:16c}. The preselected extreme months only represent themselves and have a probability of $\frac{1}{35}$ to prevent any distortion from the preselection. Alternatively, pre-selected extreme months could be added as boundary conditions to the optimization-based clustering to compute their weights endogeunsly.

\begin{figure}[h]
 \centering
  \includegraphics[scale=0.6]{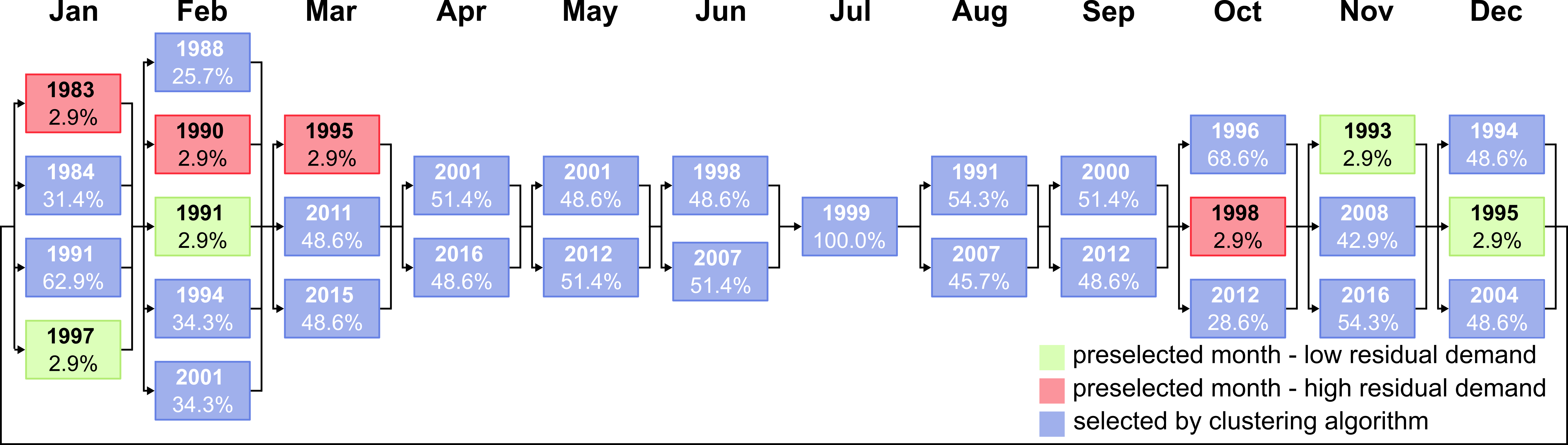}
 \caption[LoF entry]{The figure shows the result of the clustering algorithm selecting 32 representative months, including four preselected months with high residual demand and four with low residual demand. Each box represents a representative month; the first number indicates the year of the month in the original 35-year dataset, and the second number represents the corresponding probability. For instance, for April, the clustering selected the Aprils of 2001 and 2016 as representative months. The April from 1984 represents 18 of the 35 Aprils in the original dataset, which corresponds to 51.4\%.}
 \label{fig:clusterResult}
\end{figure}

Fig. \ref{fig:clusterResult} shows the result when running the prescribed algorithm to select 32 representative months, with eight preselected extreme months. Note that although we focus on examples with an equal number of scenarios per period or month, an equal number is not required for the stochastic energy model. Since wind supply and heat demand show significant interannual fluctuations during winter, while PV fluctuations in summer are minor, it is sensible that the clustering allocates more representative months to winter. Endogenously deciding the number of scenarios per period or month is another advantage of optimization-based clustering over standard methods like k-means.

\begin{figure}[h]
 \centering
  \includegraphics[scale=0.65]{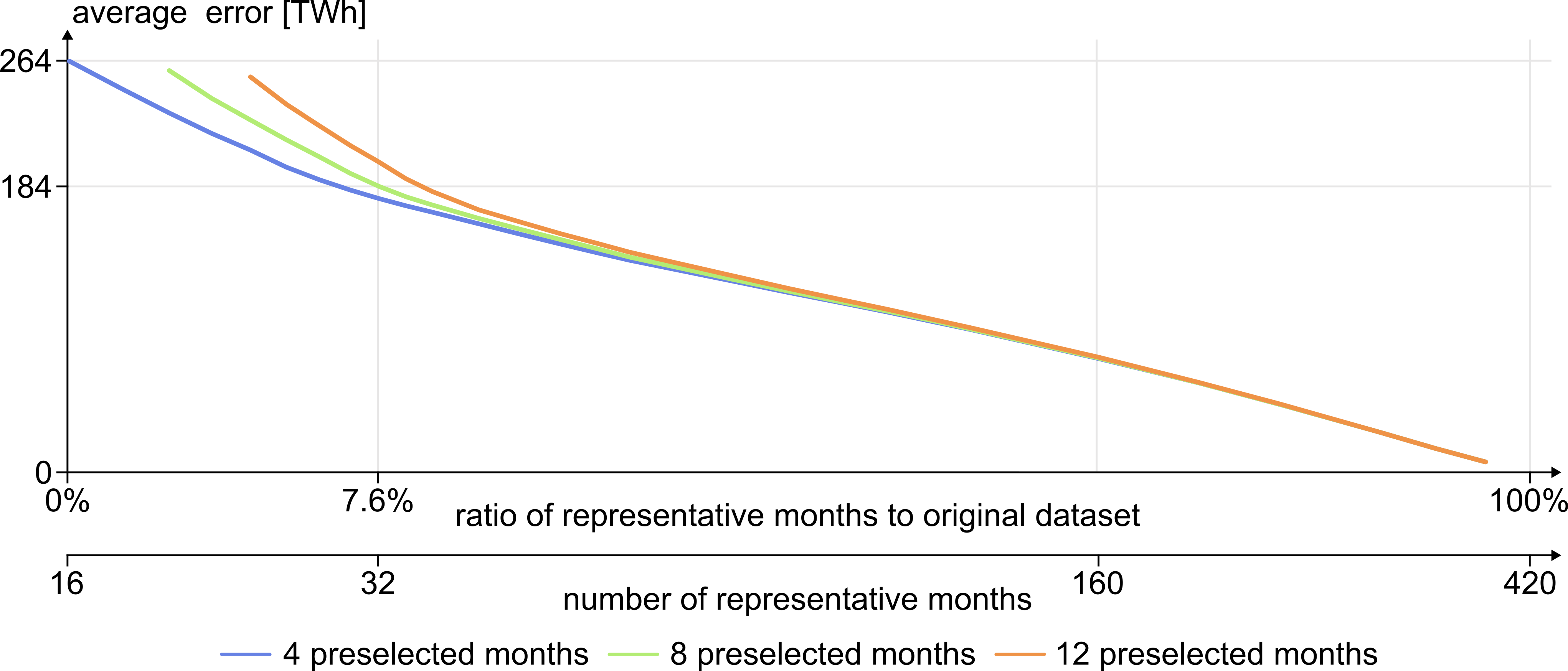}
 \caption[LoF entry]{The graph plots the average monthly error of the clustering algorithm depending on the total number of representative months and preselected extreme months. The x-axis uses a logarithmic scale.}
 \label{fig:resultClustering}
\end{figure}

Finally, Fig. \ref{fig:resultClustering} shows the average monthly error across the eight metrics, depending on the total number of representative and preselected extreme months; an error of zero implies a perfect description of the full population. Overall, preselecting extreme months comes at the expense of a higher error. The average error drops more pronouncedly until 32 representative months or 7.6\% of the total sample. Then, it gradually decreases to zero when the number of representative months approaches the total number of months in the dataset.

\subsection*{Distributed solution algorithm}

Although the introduced stochastic model and clustering algorithm massively reduce the problem size, computational tractability is still a concern for a large-scale application. Therefore, we adopt a distributed solution algorithm based on Benders decomposition to solve the stochastic optimization problem.

\begin{figure}[h]
 \centering
  \includegraphics[scale=0.65]{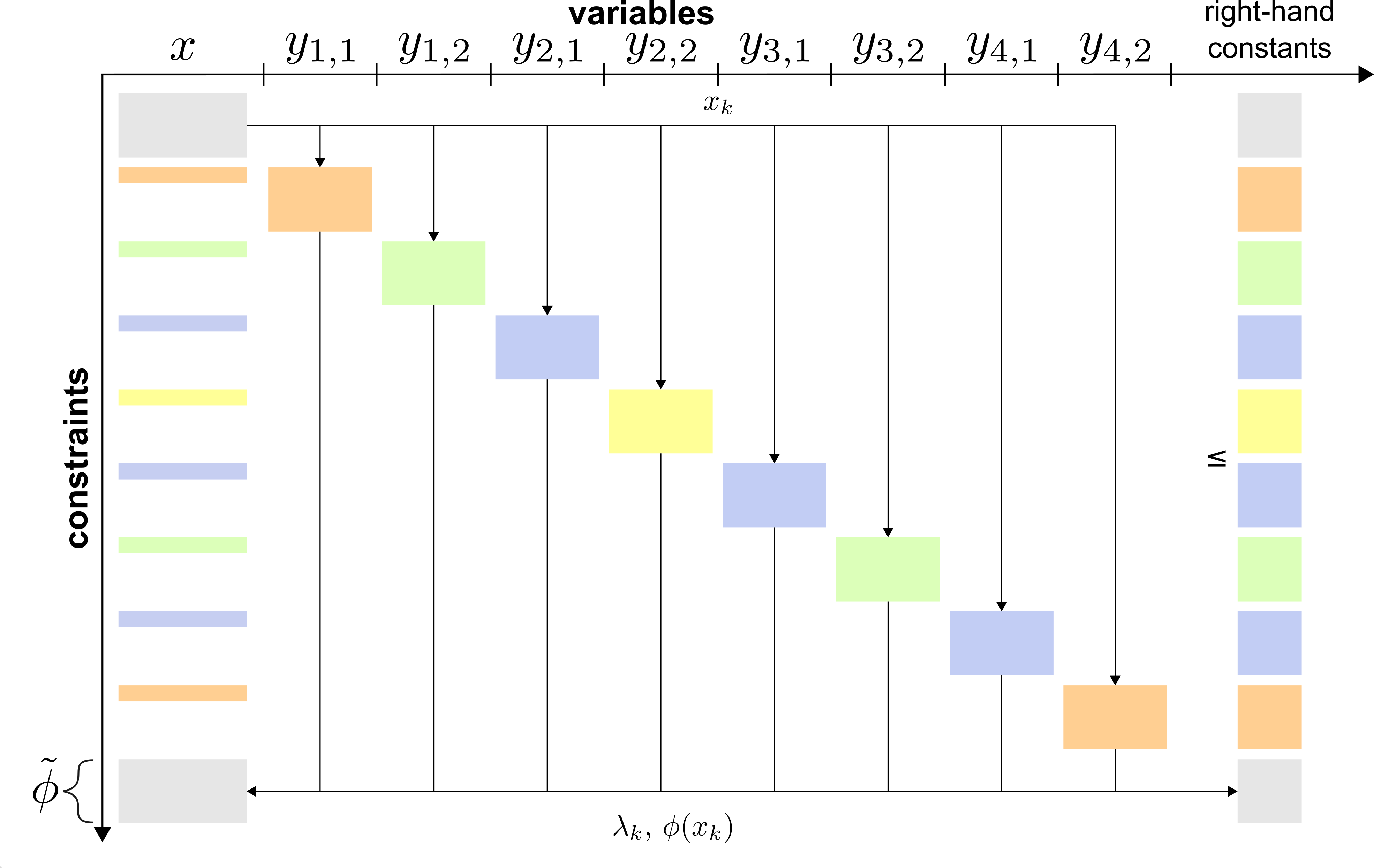}
 \caption[LoF entry]{The figure visualizes the standard decomposition of energy planning problems, based on the matrix form of the optimization problem. The top problem in grey optimizes capacity $x$ based on a linear approximation of the colored subproblems $\tilde{\phi}$. This approximation builds on solving the operational sub-problems for the proposed capacities.}
 \label{fig:benders1}
\end{figure}

The approach of Benders decomposition is to decompose the original optimization problem into smaller parts. The algorithm then iteratively determines the optimal value of the variables connecting the parts, also termed complicating variables\citep{Benders1962}. In energy planning, the complicating variables are typically capacities. Accordingly, we decompose the original problem in Fig. \ref{fig:matrix} into a first-stage top-problem that decides on capacities and into one or multiple second-stage sub-problems that decide operation, as shown in Fig. \ref{fig:benders1}. 

In iteration $k$, we first solve the top problem to obtain a capacity $x_k$ and then solve each operational sub-problem for this capacity. From the solution of each sub-problem, we take the dual variables on the capacity $\lambda_{k}$ and the objective value $\phi_{k}$ to add a linear approximation of the sub-problem at $x_k$, or a Benders cut, to the top-problem. With each iteration, this lower approximation $\tilde{\phi}$ of sub-problems in the top-problem becomes more precise, and, as a result, the solution of the top-problem converges to the optimum.

\begin{figure}[h]
 \centering
  \includegraphics[scale=0.65]{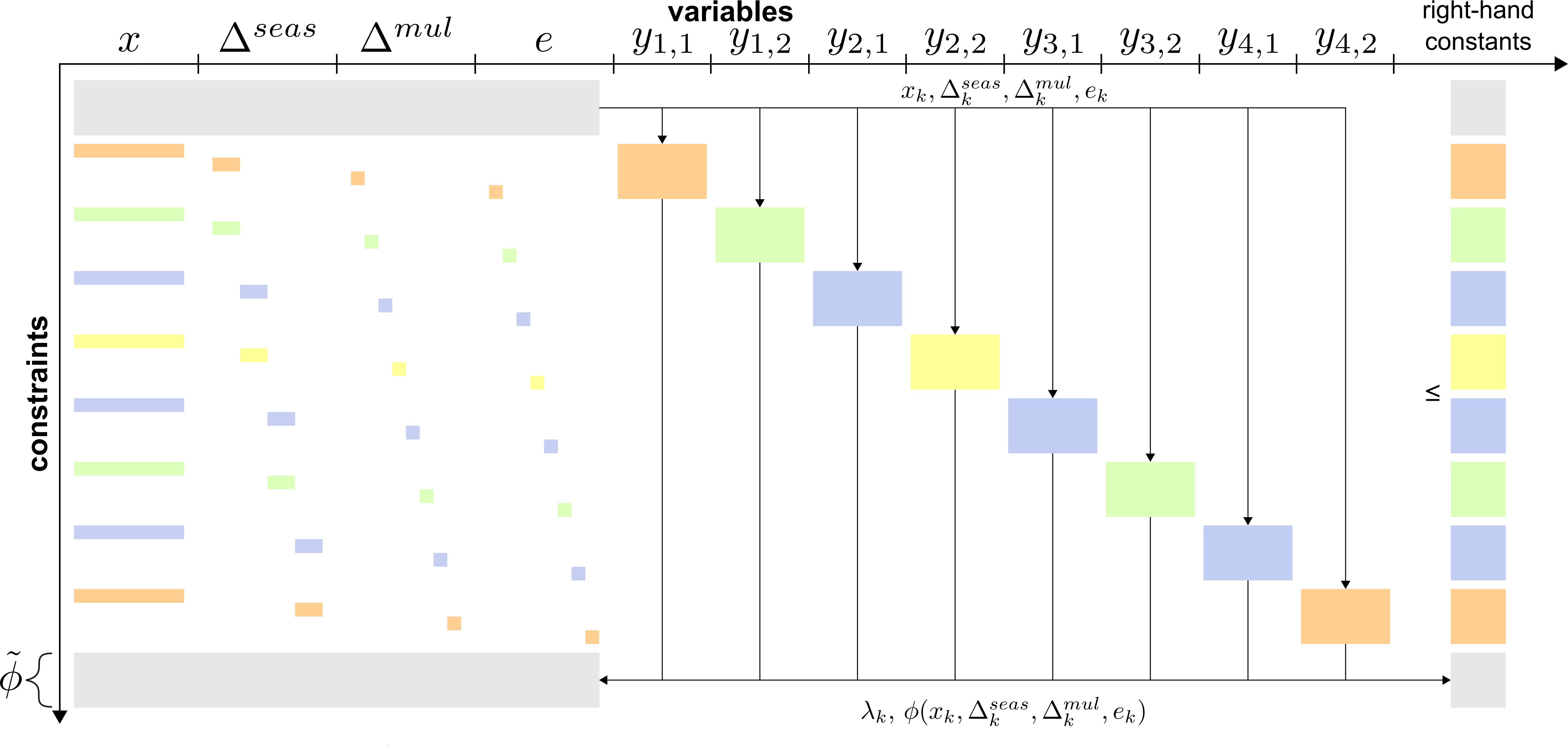}
 \caption[LoF entry]{The figure shows how the decomposition algorithm in this paper adapts the standard approach. The top problem in grey now optimizes storage levels $\Delta^{seas}$ and $\Delta^{mul}$ and emissions $e$ in addition to capacities $x$.}
 \label{fig:benders2}
\end{figure}

For the stochastic model introduced in this paper, each sub-problem corresponds to one of the stochastically independent periods---in our case, a month. To decompose the stochastic problem, we must extend the described approach to other complicating variables. First, we add the change of storage levels within the month, described by $\Delta^{seas}$ and $\Delta^{mul}$. Storage levels are complicating variables since they affect the operation but are also interrelated due to the constraints described in subsection \ref{modelImpl}. Second, we add the emissions $e$ within each month. Analogously to storage levels, the monthly emission limit affects operation, and emissions across all months are subject to a joint constraint in the top problem. In our case, the emission limit enforces a net-zero system and Fig. \ref{fig:benders2} shows the adapted algorithm. This general idea of variable splitting was previously introduced in \citet{Jacobson2024}.

Overall, adapting the decomposition adds a large number of complicating variables. Each decision on capacity expansion adds a single complicating variable, regardless of the number of months represented by a sub-problem. On the other hand, the seasonal storage component requires one complicating variable for each month in the year; the multi-year component even requires one for each represented month. In addition, we consider eight different technologies for long-term storage in each of the up to 33 regions. The complicating variables for emissions also scale with the number of represented months. However, each represented month adds a single complicating variable since there is only one global emission limit. Accordingly, decomposing the model with 32 representative months results in 2'081 complicating variables for capacities, where 1'738 relate to conversion technologies, 228 to storage technologies, and 115 to exchange infrastructure. Yet, storage levels add 3'932 complicating variables, 1'500 for the seasonal component, 2'432 for the multi-year component, and emission limits add 32. In this paper, all models were solved until an optimality gap of 0.5\%.

Due to the many complicating variables, the decomposition algorithm requires various refinements to perform well and outperform off-the-shelf solvers. Most notably, the algorithm deploys stabilization to reduce the number of iterations. Furthermore, each sub-problem runs on a different node of a computing cluster to leverage parallelization. For a detailed description, see \citet{Goeke2024}.

\newpage

%%%  The following components should appear after the 
%%%  experimental procedures (also known as methods). 
%%%  For journals using STAR Methods, these components 
%%%  should appear immediately after the discussion 
%%%  (after any "limitations" or "conclusions" subsection 
%%%  within the discussion).

\section*{Data and code availability}

The model data and script is available on GitHub: \url{https://github.com/leonardgoeke/EuSysMod/releases/tag/multiYearPaper}. 

The applied version of the AnyMOD.jl modeling framework is available here: \url{https://github.com/leonardgoeke/AnyMOD.jl/releases/tag/multiYearPaper}

\section*{Acknowledgement}

The research leading to these results has received funding from the Walter Benjamin Programm of the German Research Foundation (project number 549316936) and the ETH Zurich SPEED2ZERO initiative, which received support from the ETH-Board under the Joint Initiatives scheme. Stefano Moret acknowledges support from the Swiss National Science Foundation under Grant No. PZ00P2 202117.

Furthermore, we thank Felix Schmidt, Tom Brown, and the Department of Digital Transformation in Energy Systems at Technical University Berlin for their fruitful feedback.

\section*{Author contributions}

%%%  This component is required for most research papers. 
%%%  Mention each individual author with a statement 
%%%  outlining the contribution of each author to the work.

Conceptualization, L.G., J.W., S.M., and A.B.; Methodology, L.G., S.M., and A.B.; Investigation, L.G.; Writing – Original Draft, L.G.; Writing – Review \& Editing, L.G., J.W., S.M., and A.B.; Funding Acquisition, L.G., S.M., and A.B.; Resources, S.M. and A.B.

\section*{Declaration of interests}

%%%  This component is required for all articles, even 
%%%  if the authors have no competing interests; if 
%%%  this is the case, insert "The authors declare no 
%%%  competing interests." Please refer to the 
%%%  declaration of interests policy: 
%%%  https://www.cell.com/declaration-of-interests

A.B. served on review committees for research and development at ExxonMobil and TotalEnergies.
A.B. and S.M. have ownership interests in firms that render services to industry, some of which
may provide energy planning services. All other authors have no competing financial interests.
16

\section*{Supplemental information}

\begin{figure}[htbp]
 \centering
  \includegraphics[scale=0.65]{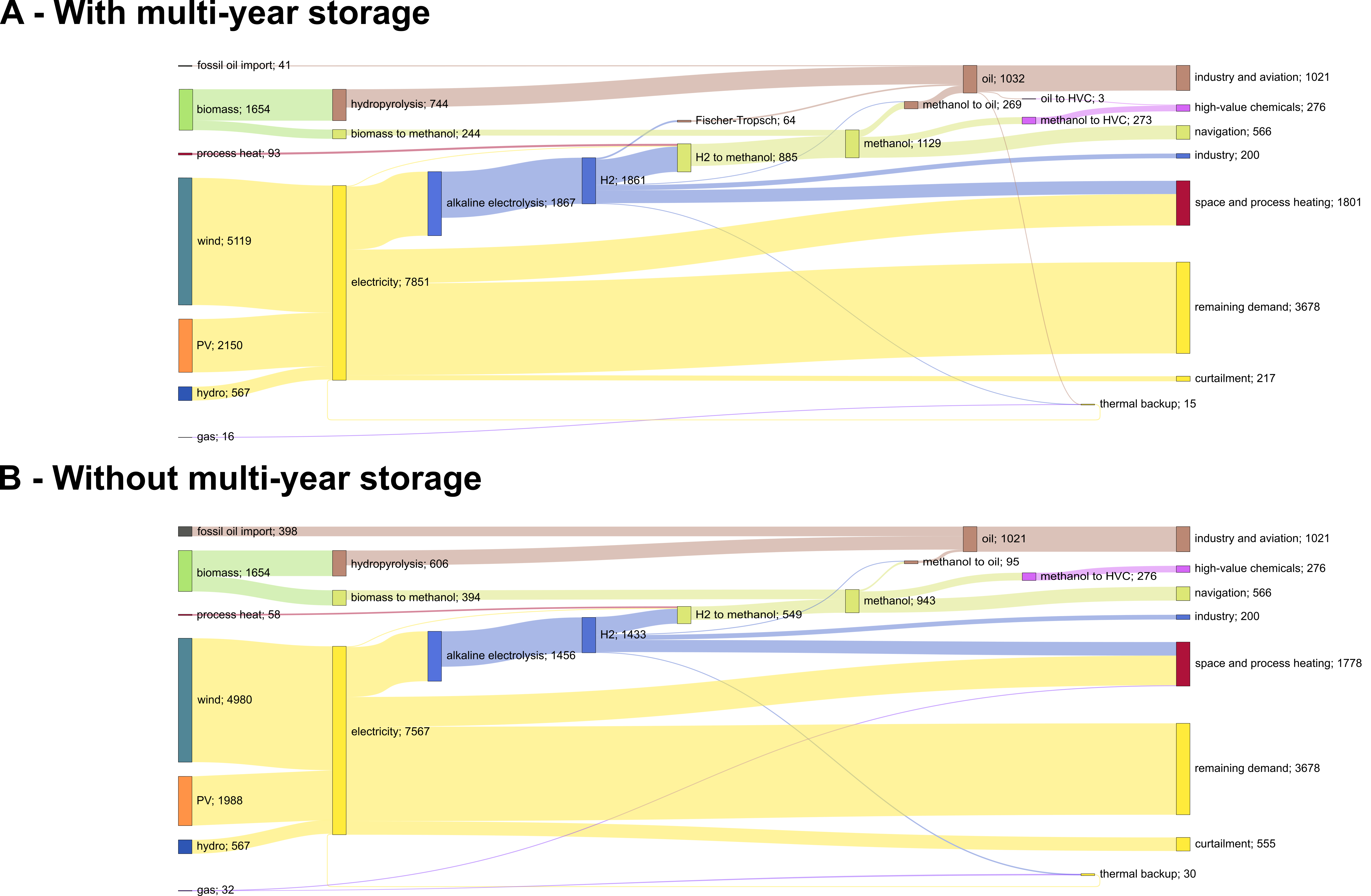}
 \caption[LoF entry]{Multi-year storage replaces fossil oil imports with renewables. The diagrams show yearly energy flow values in TWh, describing the conversion from primary energy on the left to final demand on the right. All values are expected values across the 51'840 considered climate years. The figure aggregates and simplifies the actual flows in the underlying model for clarity. For a detailed model description, see the method section. Furthermore, the appendix provides a more detailed Sankey diagram for the case with multi-year storage.}
 \label{fig:san}
\end{figure}

\begin{landscape}
\begin{figure}[p]
    \centering
    \includegraphics[scale=0.65]{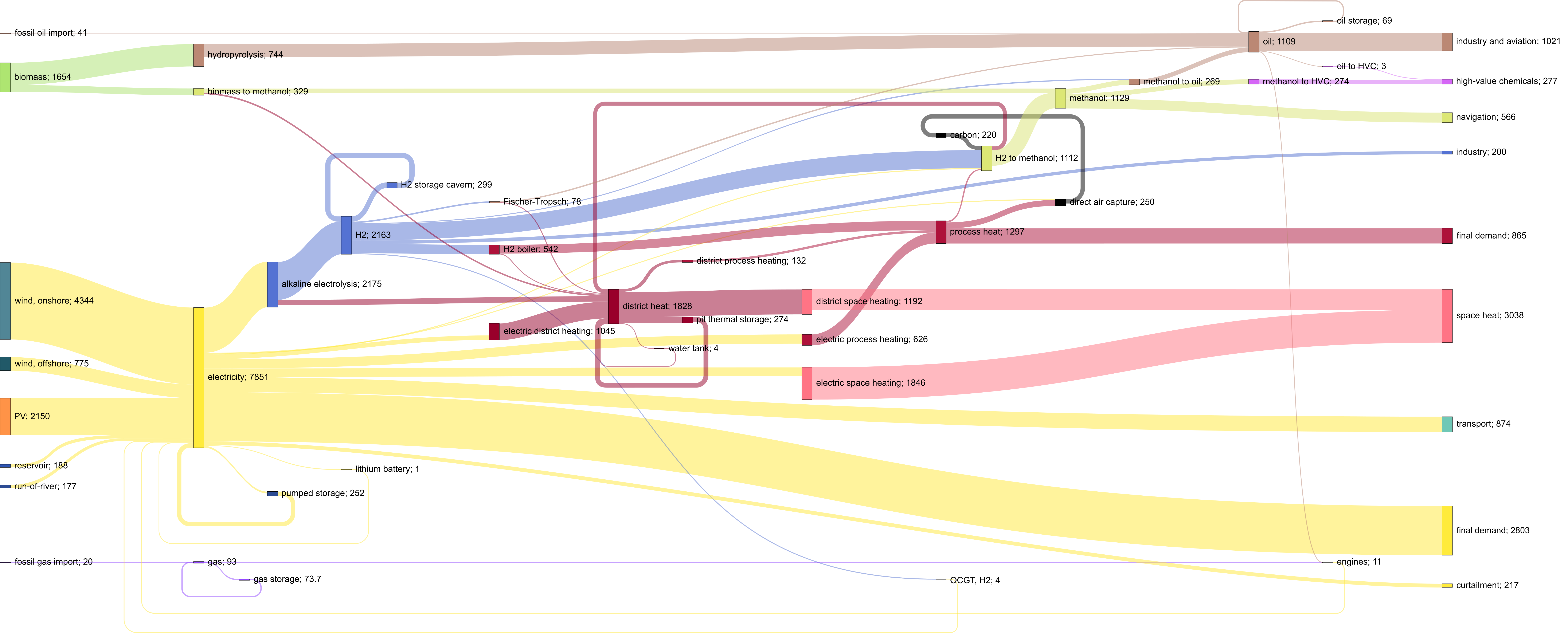}
    \caption{The diagrams show energy flows in TWh for the reference case with multi-year storage and moderate flexibility of fossil and renewable fuel imports. Most nodes in the diagrams aggregate multiple individual technologies or carriers in the model; for instance, electric space heating includes air and ground heat pumps. All numbers are aggregated across model regions.}
    \label{fig:fullSankey}
\end{figure}
\end{landscape}

%%%  Supplemental information must be uploaded as 
%%%  separate files. In the main text, please list the 
%%%  files to be included in a brief index. For details, 
%%%  please review the supplemental information guidelines: 

%%%  Journals with general "methods" or "experimental 
%%%  "procedures: 
%%%  http://www.cell.com/supplemental-information

%%%  Journals with STAR Methods: 
%%%  https://www.cell.com/STAR-supplemental-information

\bibliography{references}

\end{document}